\documentclass[preprint,12pt]{elsarticle}

\usepackage{amssymb}
\usepackage{amsmath}

\journal{Ultramicroscopy}

\begin{document}

\begin{frontmatter}



\title{Improving feature resolution and pore back effect in focused ion beam tomography of porous GaN thin films}


\author[cam,corr]{Ben Thornley}
\author[cam]{Thom R. Harris-Lee}
\author[cam]{Menno J. Kappers}
\author[cam]{Simon M. Fairclough}
\author[cam]{Rachel A. Oliver}

\affiliation[cam]{organization={Department of Materials Science and Metallurgy, University of Cambridge},
            addressline={27 Charles Babbage Road}, 
            city={Cambridge},
            postcode={CB3 0FS}, 
            state={Cambridgeshire},
            country={United Kingdom}}

\affiliation[corr]{Corresponding author - Contact: bdt28@cam.ac.uk}

\begin{abstract}

Porous gallium nitride (GaN) is a mesoporous crystalline material, typically in the form of a thin film on an unlike substrate, prepared by electrochemically etching conductive GaN. The use of porous GaN in electronics and optoelectronics is rapidly expanding, but is held back significantly by a lack of structural control and understanding of the principles of pore formation from underlying electrochemical mechanisms, where high-quality characterisation of pore morphology is essential to understanding these principles. Focused ion beam (FIB) tomography is an invaluable tool for such characterisation, but is hindered greatly by the pore back effect, where unwanted contrast appears in an image due to electrons scattering out from the back wall of a pore. No major attempts to formally quantify or assess the extent of the effect for different tomography experiments has been demonstrated. In this work, we demonstrate an advanced methodology for performing FIB tomography on porous GaN thin films, where the experiment is rotated to image pores perpendicular to the surface of the sample, and introduce new voxel intensity-based formalisms for assessing the impact of the pore back effect based on voxel intensities for individual features and the whole dataset. The new approach, which requires more complex preliminary setup, is found to significantly mitigate the pore back effect in porous GaN thin films with a range of pore morphologies without increasing the total experimental duration. The pore back effect can thus be quantified and mitigated in FIB tomography of porous GaN and similar mesoporous thin films.

\end{abstract}




\end{frontmatter}



\section{Introduction}
\label{Introduction}

Gallium nitride (GaN) is a crystalline III-V wide bandgap semiconductor that has attracted significant attention across the field of electronics and optoelectronics. In more recent years, a simple method to introduce nanoscale porosity (mesoporosity) into GaN has been developed \cite{zhang2010} and deployed in a wide range of applications, for example to enhance the piezoelectric response \cite{CALAHORRA_piezoelectricity} and to create porous GaN-based distributed Bragg reflectors \cite{ZhangDBRs}. Porosity is most often introduced through electrochemical etching (ECE), a top-down process wherein GaN films undergo selective material removal that leaves mesoporous GaN films behind. ECE takes place in simple two-electrode \cite{zhang2010} or three-electrode \cite{HarrisLee_anion} cells with conductive electrolytes, where anodic potential is applied across submerged GaN samples with typical indium contacts. The ECE experiment has a vast parameter space, where potentiodynamics \cite{HuangPVK}, electrolyte species/concentration \cite{HarrisLee_anion}, alloying concentration \cite{ABUDingan}, doping concentration \cite{Chen_HF}, amongst other tunable parameters, control the extent and morphology of porosification. Despite this, the fundamental electrochemical mechanism of ECE is elusive and poorly understood, leading to a lack of control of morphology beyond independent parameter studies. Revealing this mechanism would allow for advanced structural control that would optimise the application of porous GaN fields in many of the demonstrated applications, such as sensing \cite{RAMIZYSensor}, water-splitting \cite{BentonWaterSplitting}, composite fabrication \cite{BaiPVK}, strain relaxation \cite{KellerStrain} and photonic crystal fabrication \cite{ZhangPC}.

Effective structural characterisation of porous GaN films is an essential component in the realisation of this potential. Processing-structure relationships form a major aspect of uncovering the electrochemical processes and parameters controlling pore formation and resulting pore morphologies; structure-property relationships may only be effectively modelled and understood with quantitative analysis of pore morphology. Two-dimensional imaging modalities provide some value in the structural characterisation and can be performed very quickly - atomic force microscopy (AFM) and scanning electron microscopy (SEM) are examples of techniques which provide fast appraisal of the structure. Both of these are limited, however, as the broader three-dimensional structure of pores may be poorly represented on the top surface and along cleaved/milled cross-sections \cite{Jiawizzle}. For three-dimensional structural characterisation, and in order to reliably access any quantitative information, focused ion beam (FIB)-SEM tomography is a powerful tool.

Porous GaN films are, to some extent, an ideal use case for FIB-SEM tomography. This destructive technique is based on alternating electron-beam imaging and ion beam milling of an exposed cross-section of the material; the electron-beam images, or two-dimensional `frames' can be aligned and collated into a three-dimensional dataset by converting the image pixels into voxels (volumetric pixels), with their new additional dimension computed through tracking methods during the run. This new dimension, referred to as the milling direction, is defined relative to the orientation of the tomographic dataset, not to the GaN crystal. GaN is a sufficiently hard material that amorphisation damage from the ion beam does not become a dramatic hindrance. Additionally, since porosification of GaN in ECE requires a degree of conductivity (practically achieved by doping the material) to drive the electrochemical dissolution of material, the surviving matrix material in porous GaN films retains some moderate conductivity \cite{ZhuDBRs}, meaning imaging without extensive charging and drifting is achievable. Some three-dimensional characterisation techniques may offer advantages, whilst others may be entirely unsuitable. X-ray computed tomography (XCT) does not offer the necessary feature resolution, whilst transmission electron microscopy (TEM)-based tomography offers extremely high resolution but only across a restricted field of view. FIB-SEM tomography lies comfortably in the ideal middle ground, where pores can be well resolved across a sufficiently large field of view that hundreds or thousands of pores can be imaged at once for statistical analysis of quantitative geometric features and pore uniformity. 

Despite the high applicability of this technique in the characterisation of porous GaN, its usage is still burgeoning and rapidly improving. There are two main issues that are encountered when performing FIB-SEM tomography that restrict the ability to resolve pore features. The first is the `pore back effect' (or `shine-through artefacts') \cite{poreback} - this describes how, when imaging porous materials in SEM-based techniques, the features from the solid material that comprises the back wall of the pore are resolved in the image and contribute to the detected contrast. In standard two-dimensional imaging modalities, this does not prove to be a significant issue, but in FIB-SEM tomography this is a significant problem. Features on the back walls of pores are resolved during imaging of the cross-sectional surface at milling depths (i.e. distances along the milling direction) that are shallower than their actual position. Hence, those features contribute contrast to frames that, in the reconstructed tomograph, exist at a shallower milling depth than the true position of the feature in the sample. This leads to a `smearing' of the features in the tomograph that complicates the process of quantification and reduces the accuracy of reconstructed images extracted from the tomograph, in a way that would not be the case if opaque (i.e. not electron-transparent) features were being imaged, such as inclusions of a phase of differing composition. The second challenge with these materials is voxel anisotropy - any individual electron-beam image can have a pixel size in the sub-nm regime, but it proves difficult and unstable for the slice thickness (i.e. distance along milling direction between subsequent images) to reach such small values. Ion beam spot size and low milling speeds of low-current probes effectively restrict the lower bound of the slice thickness in the system used in this work to approximately 5\,nm. With a typical image pixel size 1\,nm and a slice thickness of 5\,nm, the resulting tomograph is essentially a three-dimensional mosaic of 1\,nm $\times$ 1\,nm $\times$ 5\,nm voxels. This anisotropy leads to a further reduction in feature resolution along the milling direction of the tomograph, which compounds with the pore back effect in porous GaN. 

The pore back effect can, to some extent, be mitigated by changing the imaging conditions of the electron beam and detector, but this offers a new set of complications. Discussion in literature of the pore back effect revolves around segmentation of datasets that are affected by the effect \cite{SALZER,JORGENSEN}, which is very pertinent in instances where it cannot be avoided. There is a lack of discussion in literature that establishes methods for ascertaining how significant the extent of the pore back effect is for a given dataset, despite the progress, particularly in recent works involving machine learning, on segmentation of affected datasets \cite{MLSegmentation}. Indeed, the most obvious way to circumvent the pore back effect is to ensure that the pore back is as far from the imaging plane as possible, so that secondary and back-scattered electrons from the back wall of the pore do not reach the detector in large numbers. Anisotropic pores are thus best imaged with the milling direction parallel to the direction of their longest dimension. For porous GaN films, ECE of GaN samples with doped layers that extend to the sample surface generally produces pore that etch downward from that surface. This study concerns the development and evaluation of a new rotated methodology for carrying out FIB-SEM tomography on porous GaN materials, and the establishment of metrics by which the severity of the pore back effect may be appraised. Traditionally, tomography has been performed using a typical set-up where trenches are milled and tracking/alignment markers are prepared on the top surface of the material. Here, we demonstrate an alternative geometry that allows for these stages to instead take place on a cleaved and prepared cross-sectional edge of the samples, allowing for imaging to take place with the milling direction perpendicular to the material surface and therefore more aligned to the long dimension of the pores. The new technique is appraised by comparing the feature resolution of the conventional approach to FIB-SEM tomography with the new approach, by capturing tomographic datasets on three porous GaN samples in both setups. 

\section{Methods}
\label{Methods}

\subsection{Sample Fabrication}
\label{fabrication}

\begin{figure}
    \centering
    \includegraphics[width=1\linewidth]{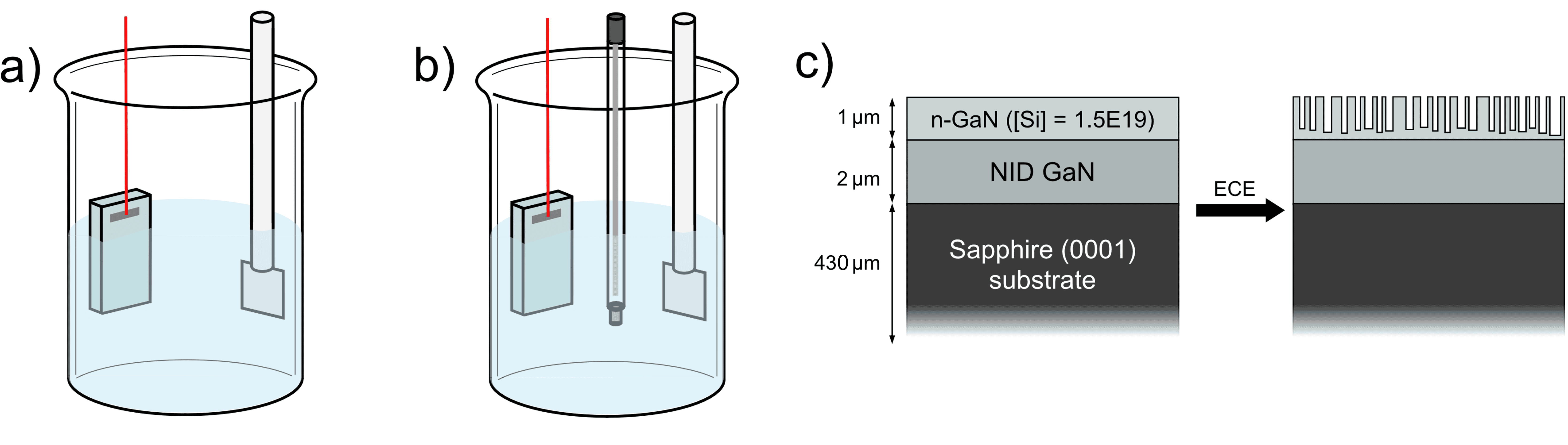}
    \caption{Schematic of porosification process showing a) 2-electrode ECE, b) 3-electrode ECE and c) GaN sample structure showing conductivity-selective porosification}
    \label{fig1}
\end{figure}

Three porous GaN samples were used in this work, each prepared by ECE of samples from identical wafers. Different ECE conditions were selected for the three samples to produce drastically different pore morphologies, though each used the same etching apparatus. Wafers were prepared by metal-organic vapour phase epitaxy using an Aixtron close-coupled showerhead reactor. The wafers consisted of commercial 2-inch diameter sapphire c-plane substrates, onto which a non-intentionally doped (NID) GaN buffer of thickness 2.4\,\textmu m was deposited, onto which a further 1\,\textmu m of Si-doped n-GaN ([Si] = $1.5 \times 10^{19}$ cm$^{-3}$) was further deposited. All electrochemical etching (ECE) was performed using a Gamry Reference 3000 potentiostat. Cleaved wafer pieces were soldered with metallic indium contacts onto which wired connections could be directly added. Sample 1 and Sample 2 were etched in a three-electrode cell configuration, where the sample, a platinum plate counter electrode (Osilla, 10\,mm $\times$ 10\,mm $\times$ 0.1\,mm) and a 3\,M Ag/AgCl reference electrode were all partially submerged into electrolyte solution. Sample 3 was etched in a two-electrode cell, where the same conditions as above applied but the reference electrode was excluded. In all instances, the potentiostat applied a constant potential across the GaN samples and the counter electrode, and recorded the current that flowed between them as etching proceeded. The end-point of porosification was indicated by current decreasing to low, relatively stable levels ($\approx$ 10\,\textmu A). Porosification of GaN is known to be a conductivity-selective process, such that the extent of doping determines the extent of porosity for a given ECE condition. At low background currents, therefore, the n-GaN layers had undergone complete porosification whilst the NID GaN layers remained unaffected, leaving all three samples to consist of a 1\,\textmu m porous GaN layer on an NID GaN buffer, each with a different pore morphology.

For the three samples, the voltage, etchant species/concentration and use of two-/three-electrode cells were changed to give the different pore morphologies. Sample 1 was etched in a 0.1\,M aqueous solution of sodium carbonate (Na$_2$CO$_3$, Afla Aesar, 98\%) at a potential of 8\,V in a 3-electrode setup. Sample 2 was etched in a 0.1\,M aqueous solution of sodium Na$_2$CO$_3$ but at a potential of 4\,V in a 3-electrode setup. Sample 3 was etched in a 0.25\,M aqueous solution of oxalic acid (H$_2$C$_2$O$_4$, Sigma-Aldrich, 99.97\%) at a potential of 8\,V in a 2-electrode setup. These result in three very different pore morphologies, so that the two approaches to tomography may be compared for different porous structures (this is discussed further in section \ref{evaluation}). All samples were etched in ambient light and temperature conditions.

Aside from characterisation in FIB-SEM, all samples were also characterised by atomic force microscopy (AFM) in a Bruker Dimension Icon set to PeakForce tapping mode (using Bruker ScanAsyst HPI tips) to appraise the surface morphology. Datasets were processed and exported using Gwyddion software \cite{Necas2012}.

\subsection{Focused Ion Beam Tomography}
\label{FIB tomo}

All SEM and FIB-SEM in this work was performed using a Zeiss Crossbeam 540 FIB-SEM with a Zeiss Gemini 2 column installed. To ensure fidelity in comparing across various datasets, all imaging in tomography was performed using an electron accelerating voltage of 2\,kV and an electron probe current of 65\,pA. We note that the Ga$^{3+}$ ion beam accelerating voltage was also held constant throughout this work, at 30\,kV, but probe current was varied substantially for different parts of tomography setup. 

\begin{figure}
    \centering
    \includegraphics[width=1\linewidth]{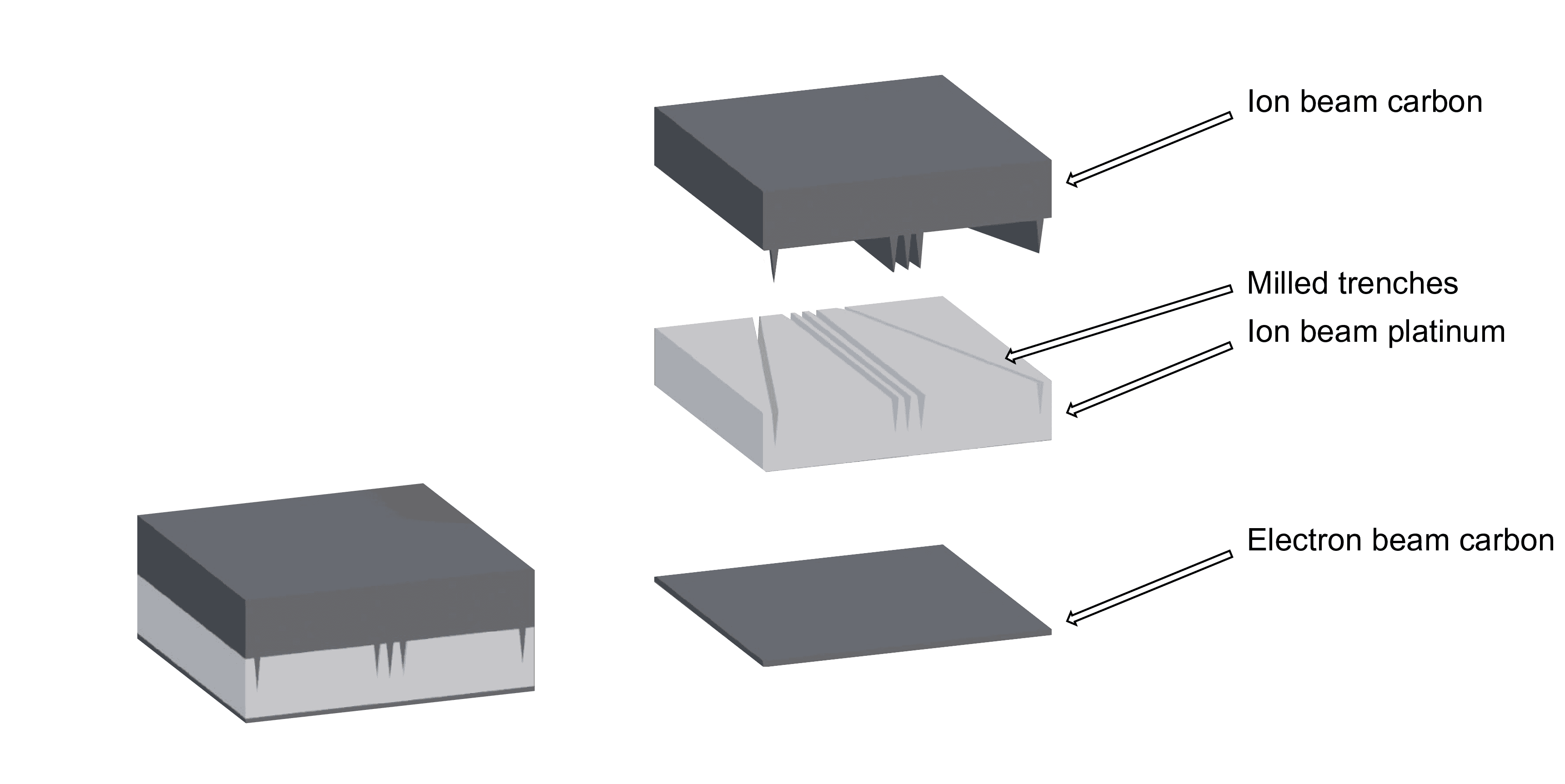}
    \caption{Tracking structure needed for tomography acquisition, shown as (left) real spacing and (right) exaggerated spacing. The tracking structure consists of initial electron beam-deposited carbon, overlaid with milled platinum and backfilled/overlaid carbon}
    \label{fig2}
\end{figure}

Tomography datasets were acquired using the Nanotomography plugin package of the Zeiss Altas Engine/Atlas5 software. Tomography runs proceeded using `continuous milling and imaging' mode, where the ion beam milling and electron beam imaging take place on the cross-sectional face of the porous GaN simultaneously, to aid in charge neutralisation and prevent drifting. Tomography runs in all instances require the preparation of a Pt/C tracking structure (shown in Figure \ref{fig2}) laid onto the region of interest (ROI) to aid in alignment and tracking during the tomography run. This consists of an initial carbon layer of approximately 100\,nm thickness prepared by electron beam-assisted deposition, then an approximately 1\,\textmu m thick Pt pad, into which five straight tracking lines are milled. Three of these run parallel down the centre of the pad, whilst the other two are on the outside of these and converge inwards. These lines are backfilled with carbon and a further 1\,\textmu m thick carbon pad is overlaid. With the tomograph ROI overlaid with this structure, trenches are then milled into the surrounding porous GaN to prevent issues with redeposition and ensure good clearance for electron beam imaging. Finally, and crucially, the ion beam is used to mill a trench that cuts into the face of the deposited structure. After the carbon is deposited onto the platinum, the surface of the tracking structure is not flat, and the carbon tends to coat the side walls of the platinum. Thus, after deposition only, the five tracking lines are not clearly visible and the tomography run cannot begin. Hence, the front edge of the tracking structure must be partially milled into before the tomography run can begin. This is conventionally done as part of the preparation of a trench at the front of the tracking structure, which also exposes the features in the region of interest. Tracking marks are then visible on the ion beam polished face of the tracking structure as five visible V-shaped depressions in the Pt/C interface.

Tomography runs proceeded by milling and imaging cross-sectional faces of the porous GaN, imaging the five tracking marks and region of interest for all frames. After the completion of runs, datasets were aligned using the three parallel tracking marks as a reference feature and were registered by utilising the outer tracking marks to infer the specific slice thickness of each frame. The distance between the outer tracking marks, combined with the known angle at which they are milled, allows for the individual thickness of material removed to be known for each slice. The frames could then be collated into voxels with their individual computed slice thicknesses and be exported as full tomographic datasets (`tomographs'). All imaging processing and image reconstruction was performed using Dragonfly 3D World software. 

\subsubsection{Cross-sectional tomography}
\label{cross sectional tomo}

\begin{figure}
    \centering
    \includegraphics[width=0.9\linewidth]{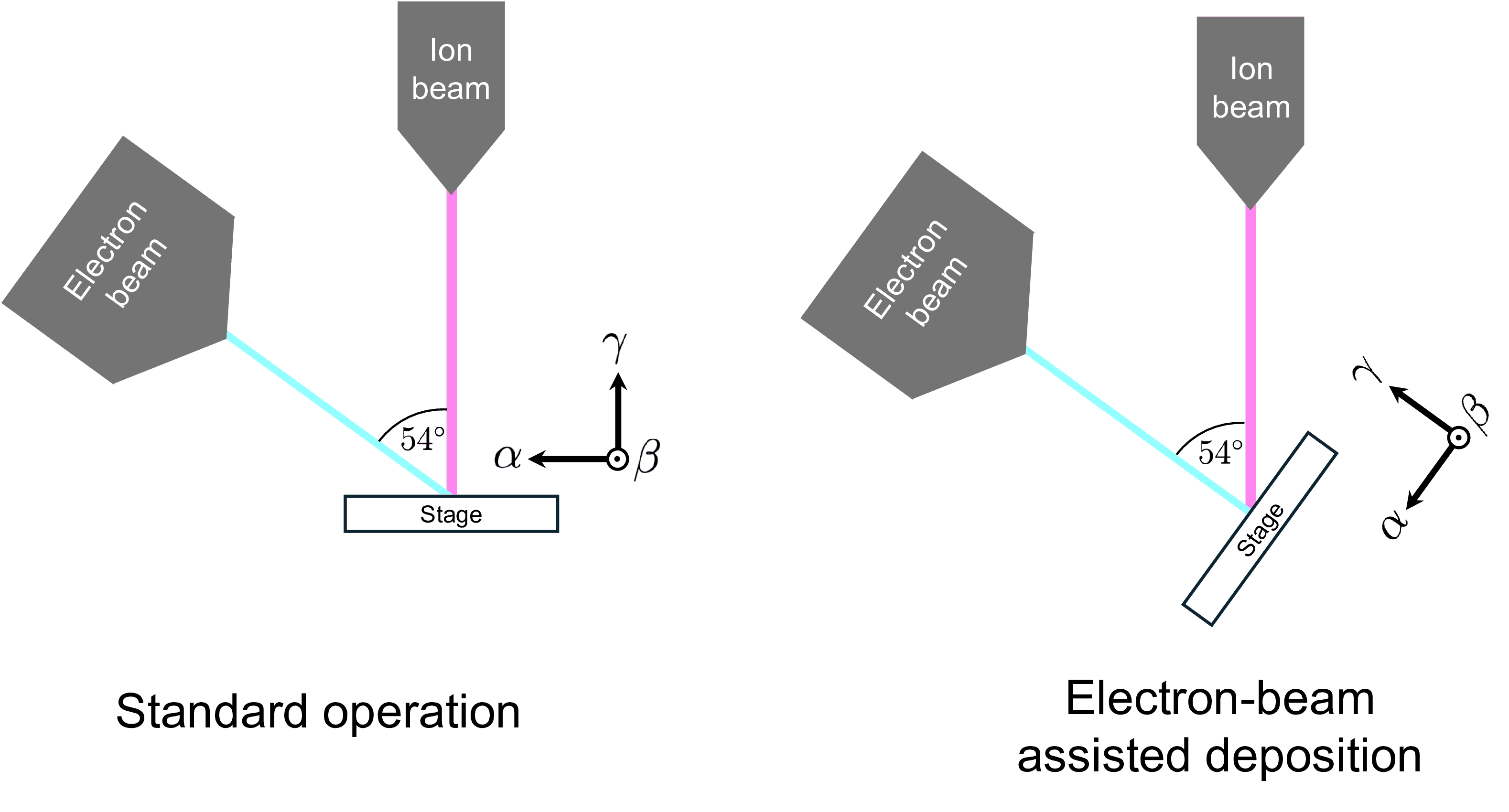}
    \caption{Schematic diagram showing the orientation of defined $\alpha$, $\beta$ and $\gamma$ axes relative to the ion/electron beams and the sample stage, during standard operation and with the stage tilted to perform electron beam-assisted deposition}
    \label{fig2.2}
\end{figure}

It is useful for understanding the geometry of the setup and operation steps of tomography experiment to define a reference set of orthogonal axes which remain constant. This work compares different orientations of capturing tomographs, so these axes are defined relative to the instrument, and the orientation of the sample throughout is defined relative to the axes. Axes are denoted $\alpha$, $\beta$ and $\gamma$. The $\gamma$ axis is defined parallel to the ion beam, as shown in Figure \ref{fig2.2}. The electron beam and ion beam have a fixed angle of 54$^\circ$ between them, from instrumental installation. The ion beam is thus perpendicular to the stage during operation, so the $\alpha$ and $\beta$ are directions which lie in the plane parallel to the stage. These are distinguished as the $\beta$ axis is perpendicular to both beams, and the $\alpha$ axis is not. Each time a new slice is cut from the surface during tomography, so that a new image may be captured, the ion beam must displace along the (negative) $\alpha$ axis, such that the milling direction is always (anti)parallel to the $\alpha$ axis. The above is true for all setup and operation except for during electron-beam assisted deposition, where the electron beam is now parallel to the $\gamma$ axis, and $\alpha$ and $\beta$ remain in the plane parallel to the stage. Since this can be achieved by rotating the stage, the distinction is clarified here, but rotating the stage to access these different operations will not be shown in schematic figures throughout.

Throughout this work, `cross-sectional tomography' refers to experiments in the simpler and more conventional approach. All previous work that we have published has used this simpler approach \cite{THORNLEY2026121957,Jiawizzle}. In brief, during this setup, porous GaN samples are mounted flat onto SEM stubs, so that the top surface of the porous GaN is perpendicular to the ion beam and $\gamma$ axis during all tomography setup and runs, and that the electron beam, during the experiments, captures images of the cross-sectional face of the porous GaN, with a pre-applied tilt correction accounting for the non-orthogonal incidence of the imaging beam and milled surface. These frames thus show the full length of pores from the top surface vertically down to the bottom of the porous layer. Here, the milling direction/$\alpha$ axis of the tomograph thus lies in the plane parallel to the material surface.

\begin{figure}
    \centering
    \includegraphics[width=1\linewidth]{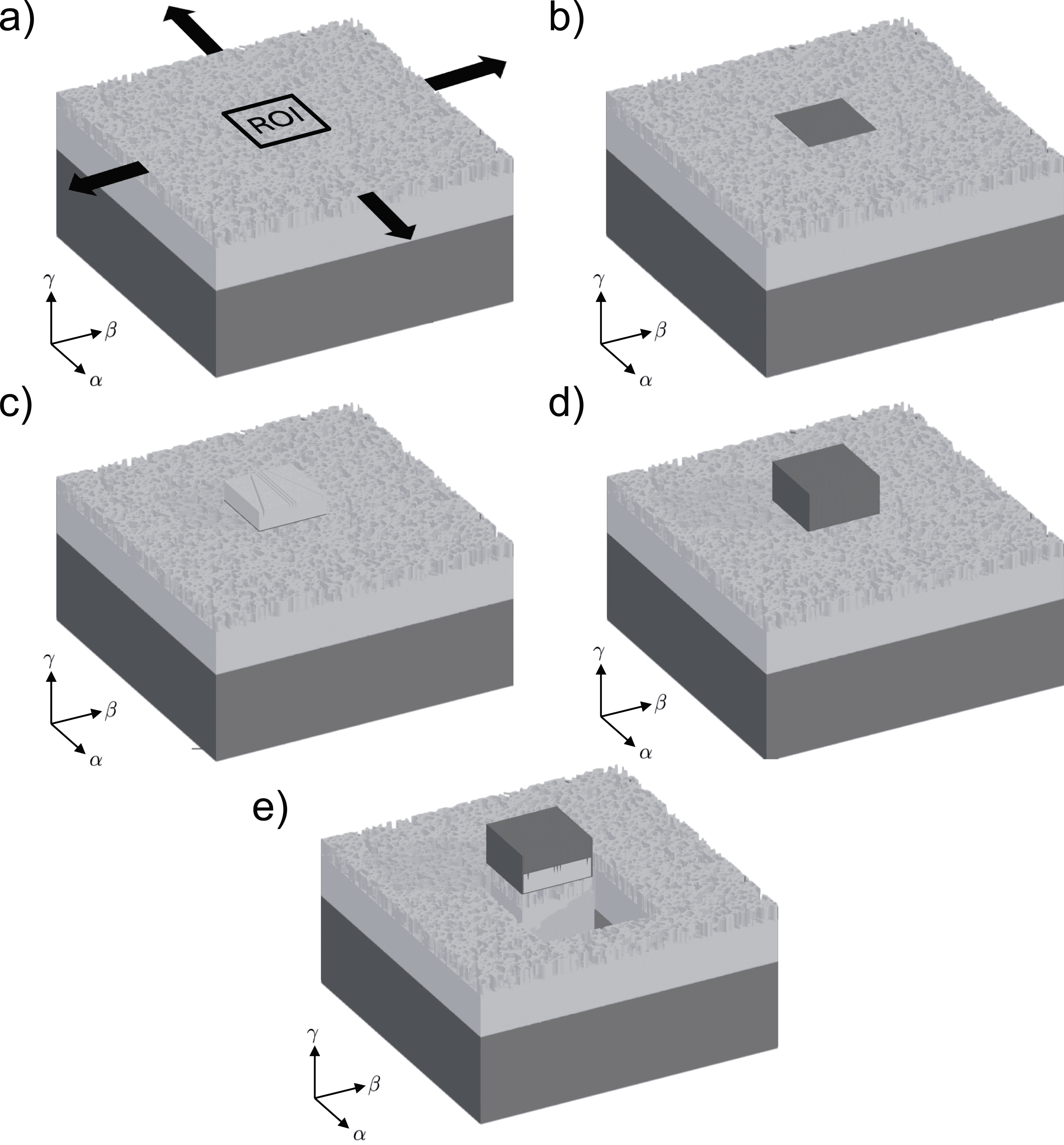}
    \caption{Schematic illustration of setup for cross-sectional tomography, showing a) the region of interest positioned in the centre of the sample (black arrows indicate extension of the sample beyond the image); b) initial carbon deposition onto the region of interest; c) ion beam deposition and platinum with subsequent milled tracking marks; d) overlaid carbon which coats the sides of the platinum and e) the milling of trenches to reveal tracking marks}
    \label{fig3}
\end{figure}

\begin{figure}
    \centering
    \includegraphics[width=0.8\linewidth]{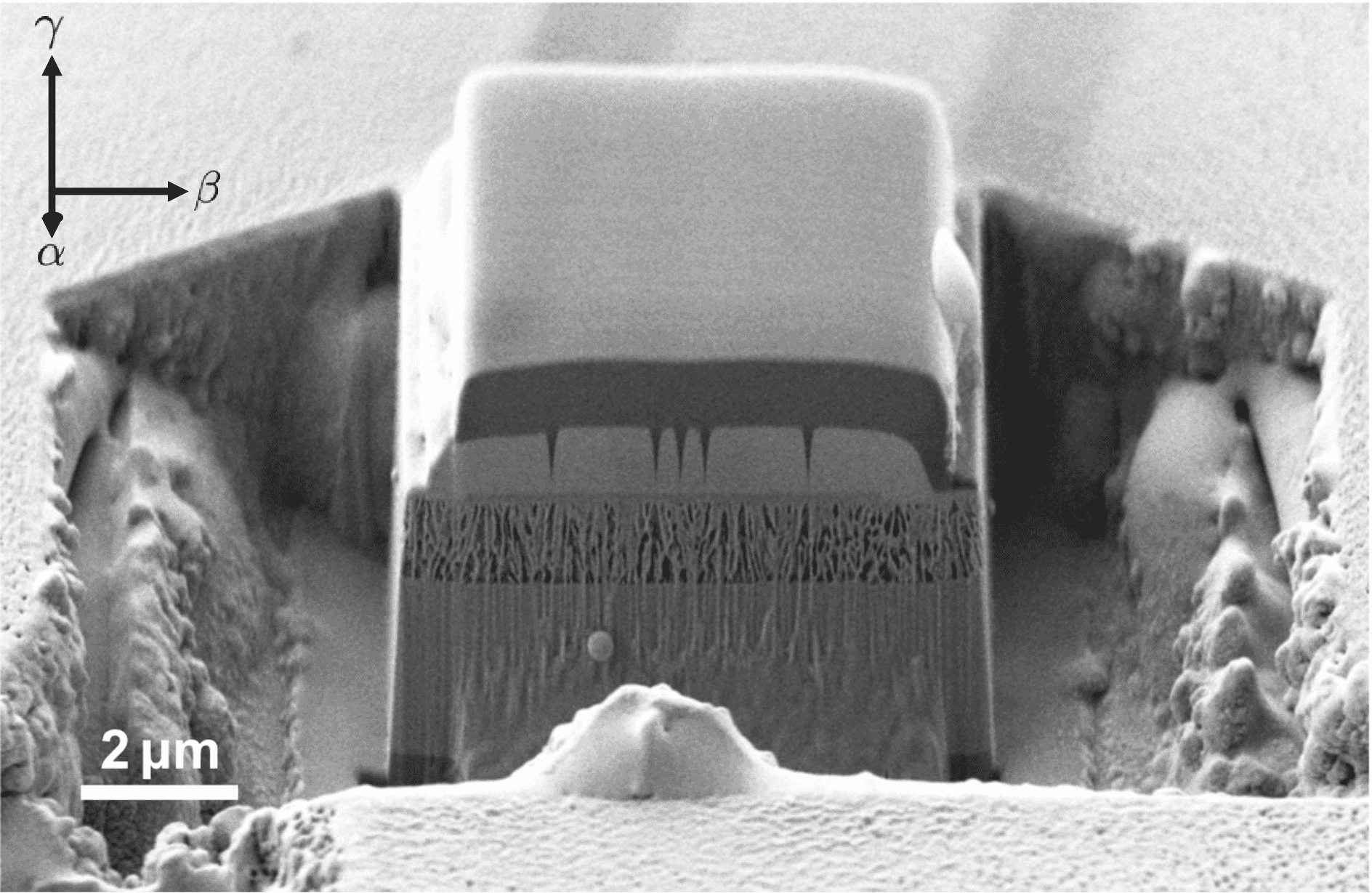}
    \caption{SE image of porous GaN after setup stages for cross-sectional tomography}
    \label{fig4}
\end{figure}

To prepare the three samples for tomography in this setup, each was mounted onto a flat stub, and the ROI was selected arbitrarily in the centre of the sample (as shown in Figure \ref{fig3}a). To protect the surface from ion beam damage, the chosen ROI was first overlaid with electron beam-assisted carbon deposition, where the built-in gas injection system (GIS) and a high electron probe current of 7500\,pA were utilised, depositing approximately 100\,nm of carbon onto the surface (shown in Figure \ref{fig3}b). The same GIS was then used for ion beam deposition of Pt to produce the 1\,\textmu m thick Pt pad onto this carbon layer, with a moderate ion beam current of 300\,pA or 700\,pA. Into this pad, the tracking lines were milled using a 20\,pA ion beam current, and backfilled with carbon with the same current. The platinum pad with the milled tracking lines is shown in Figure \ref{fig3}c. The whole structure was then overlaid with 1\,\textmu m of carbon using the 300\,pA ion beam current, as shown in Figure \ref{fig3}d. As mentioned previously, when the carbon is deposited, it tends to cover the sides of the platinum pad slightly; hence only carbon is visible after deposition in Figure \ref{fig3}d. Trenches were milled into the surrounding GaN for clearance. Finally, the deposited structure was milled with the 300\,pA ion beam current probe to remove the edge and expose the 5 tracking marks. The tracking structure after trench milling is shown in Figure \ref{fig3}e, where the tracking lines appear as five features in the tracking structure. Tomography runs then proceeded as standard, where the ion beam could mill the exposed surface and the electron beam could image the five tracking marks and the region of interest, with the same orientation of the GaN with respect to the instrument axes kept constant. A real image of the porous GaN after this setup is provided in Figure \ref{fig4}, comparable to the equivalent step in the schematic shown in Figure \ref{fig3}e.

\subsubsection{Plan-view tomography}
\label{plan-view tomo}

\begin{figure}
    \centering
    \includegraphics[width=0.9\linewidth]{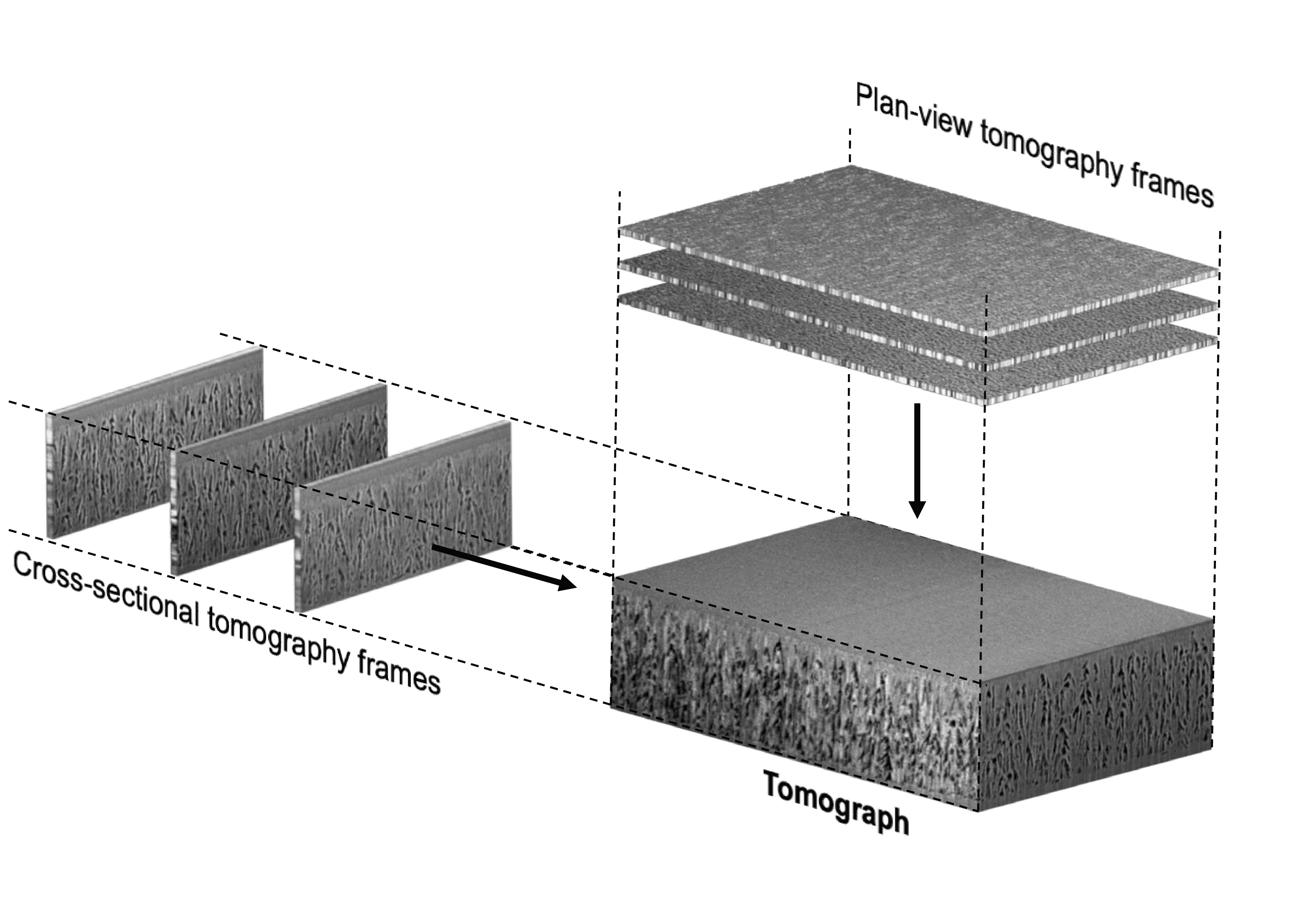}
    \caption{Schematic diagram of the orientation of imaging inputs for tomography of porous GaN in cross-sectional (conventional) and plan-view (surface-plane) modes}
    \label{fig5}
\end{figure}

This work features an alternative setup for FIB-SEM tomography of these materials, referred to throughout (in order to draw distinction from the cross-sectional tomography mentioned previously) as `plan-view tomography'. The principle of this is to rotate the entire tomography experiment by 90$^\circ$ such that the electron beam is no longer imaging the whole depth of the porous layer but instead is imaging the exposed surface of the porous GaN layer. This distinction is illustrated in Figure \ref{fig5}, where the orientation of the images that comprise the tomograph are shown for each imaging mode relative to the 1\,\textmu m porous layer. In the plan-view experiment, the ion beam mills material to slice the ROI parallel to the plane of the surface, and each frame, rather than featuring the entire porous layer in the cross-sectional perspective, shows the in-plane pore morphology of the sample at a given depth through the porous layer. The milling direction/$\alpha$ axis is thus rotated relative to the sample - now, the milling direction is perpendicular to the surface of the GaN. To rotate the axes relative to the GaN, this obviously requires mounting the sample at a different angle. Use of angled SEM stubs thus becomes essential, but, despite the simple concept, preparation and execution of this setup proved to be non-trivial.

\begin{figure}
    \centering
    \includegraphics[width=0.91\linewidth]{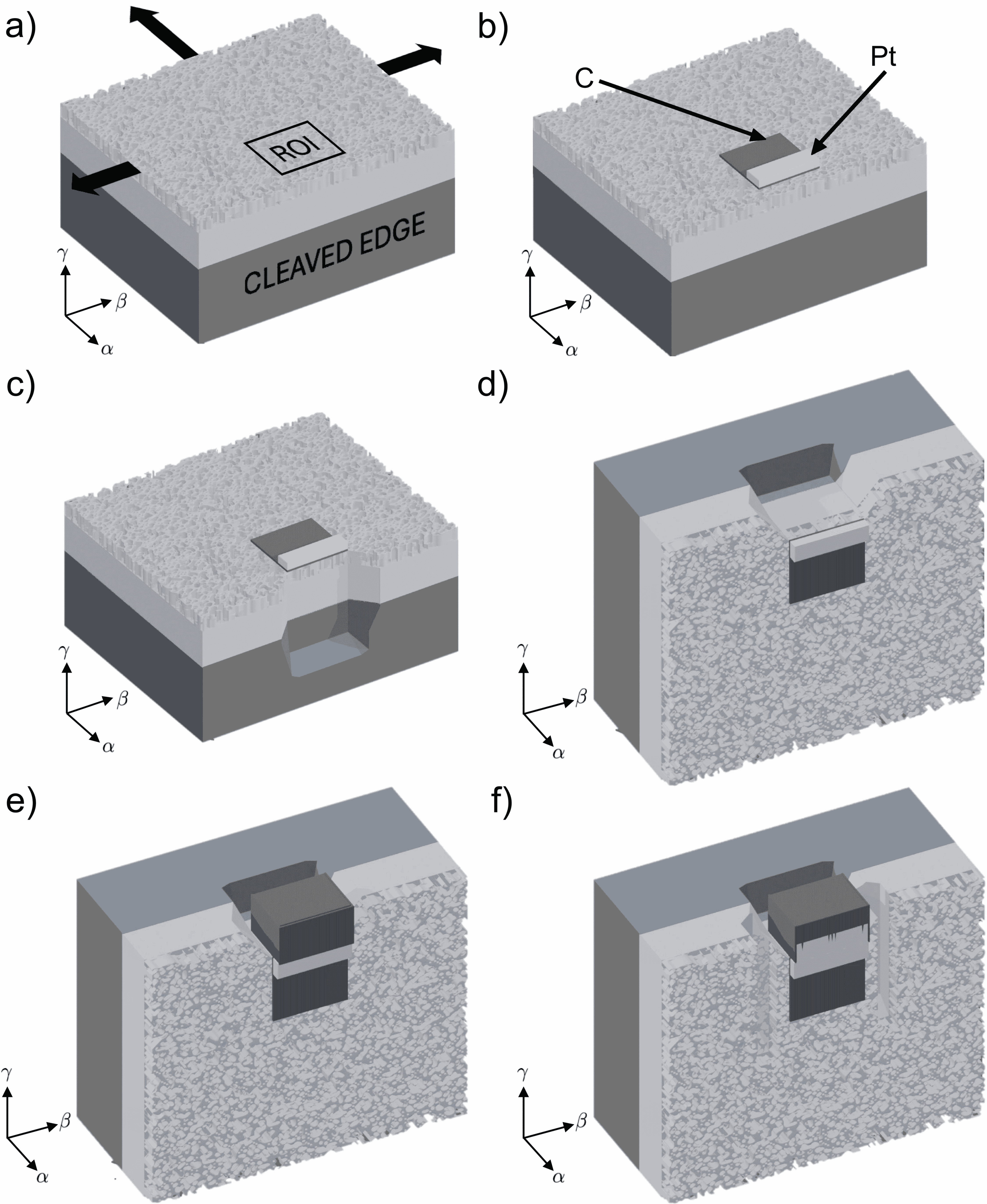}
    \caption{Schematic illustration of setup for plan-view tomography, showing a) the region of interest positioned near to the cleaved edge of the sample (black arrows indicate extension of the sample beyond the image); b) initial carbon deposition onto the region of interest and platinum deposition to form shelf; c) milling of trenches to partially remove shelf; d) rotation of the sample to a cross-sectional stub, so that the shelf is horizontal and the surface over the region of interest has been planarised; e) deposition of tracking structure extending out onto shelf and f) milling of side trenches and frontal trench to reveal tracking marks}
    \label{fig7}
\end{figure}

Development of this experiment was thorough. Discussion of the major practical challenges that arose between conceptualising the experiment and it being realised are provided, for the interested reader, in the supplementary material (section 2). The current methodology has been developed to circumvent the issues outlined there. It is also worth noting that the total range of the m-axis of the stage in the Crossbeam 540 instrument presented another limitation, as using standard 90-degree stubs placed the sample too far away from the stage for eucentric height to be achieved at the limit of the m-axis. Hence, plan-view tomography must be performed on small pieces of material using short 90$^\circ$ stubs in the instrument used in this work. The setup used in this work consists as follows: the etched GaN samples were cleaved down to a small size, then partially coated with silver paste onto flat SEM stubs. Initial stages thus took place with the same orientation as in cross-sectional tomography, where the $\gamma$ axis is perpendicular to the sample surface. The edges of the pieces with the fracture surfaces that visually appeared to be the most straight were left uncoated and, to ensure that suitable positions could be located, both cleaved pieces of GaN were mounted at once with the edges formed by either side of the same fracture surface left uncoated. 

A suitable ROI was then selected at a distance of approximately 1\,\textmu m from the cleaved edge, at a position where the cleaved edge itself is relatively linear and the sapphire substrate is not visible in plan-view imaging (i.e. there is no significant ledge at the GaN/sapphire interface, and the sapphire has either cleaved with a fracture surface well aligned with that of the GaN, or the porous GaN/NID GaN is overhanging). This is shown in Figure \ref{fig7}a. Onto the region of interest, protective carbon was again deposited via electron beam-assisted deposition with a probe current of 7500\,pA. For this work these were generally large, approximately 9\,\textmu m $\times$ 6\,\textmu m, parallel to the cleaved edge and with approximately a 1\,\textmu m clearance away from it. Onto this protective carbon, a 1\,\textmu m-thick platinum pad, with a length equal to the carbon pad and a width of 2\,\textmu m, was deposited at the edge of the carbon pad, closest to the cleaved edge, as shown in Figure \ref{fig7}b. This ion beam assisted deposition used the 300\,pA ion beam, and the protective carbon primarily ensures that the surface of the porous GaN ROI was not damaged during this ion beam exposure. Finally, the same 300\,pA beam was used to mill a trench that extended from the cleaved edge partially into the platinum pad. This leaves behind an exposed porous GaN surface with a flat platinum feature that terminates at the same exposed face. This is shown in Figure \ref{fig7}c, where the milling extends from the cleaved surface inwards reaching a small distance into the platinum pad.

After this stage, prepared samples were unmounted from the flat stubs and mounted into 90$^\circ$ angled-edge stubs, such that the prepared edge faced upwards out of the stub and was parallel with the flat surface of the sample holder. Hence, the sample is rotated relative to the $\alpha/\beta/\gamma$ axes (specifically by -90$^\circ$ around the $\beta$ axis and 180$^\circ$ around the $\gamma$ axis), and this rotation is illustrated in the change in the orientation of the sample between Figure \ref{fig7}c and \ref{fig7}d; no additional setup takes place between Figure \ref{fig7}c and \ref{fig7}d, but the remounting of the sample places the prepared platinum feature at a new orientation to the ion beam/$\gamma$ axis. The same region of interest could now be found where, due to the previous stages, the cross-sectional edge of the prepared surface, now perpendicular to the $\gamma$ axis, is suitably flat for further setup. The milling of the trench ensures that the prepared GaN surface is planar enough for deposition, whilst the platinum shelf prevents the surface of the GaN from being exposed to the ion beam during deposition, so that no additional damage takes place, and the new platinum shelf can be used to extend the deposited tracking structure out beyond the porous GaN surface, such that there is now clearance with which the structure can be partially milled to expose the tracking marks and the tomography run can take place without any undesirable preceding milling of the GaN surface in the ROI. 

\begin{figure}
    \centering
    \includegraphics[width=0.8\linewidth]{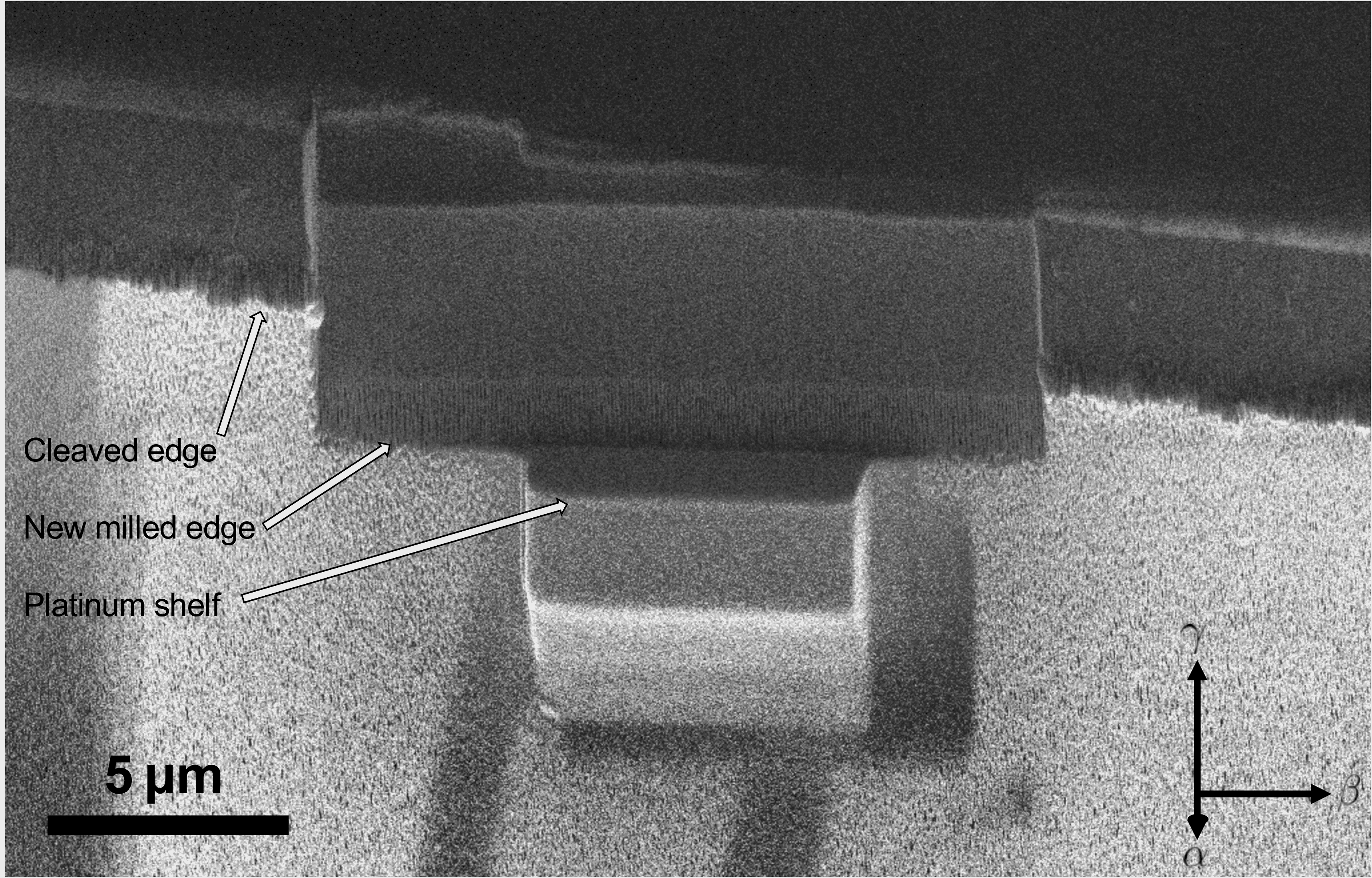}
    \caption{Labelled SE image of prepared shelf for plan-view tomography setup}
    \label{fig8}
\end{figure}

Setup can then take place essentially the same as before, onto the cross-sectional surface prepared previously. The initial carbon and further platinum pads are deposited not only onto the porous GaN, but extending further outwards onto the platinum pad, and further inwards into the NID GaN such that the entire porous layer is inside the ROI (shown in Figure \ref{fig7}e). Further milling of tracking marks and deposition on top with carbon take place as previously, with the only difference being that material does not need to be removed from in front of the ROI for the electron beam clearance since the process takes place on the edge, so there is no material available to be removed. Once the edge of the deposited structure has been removed, by milling away much of the shelf, the deposited tracking structure is left overhanging slightly over the porous GaN edge, as shown in Figure \ref{fig7}f. Since the tracking marks are exposed, the tomography run can commence. The features imaged by the electron beam for the first section of the run, therefore, just consist of the overlaid carbon onto the porous GaN surface from the start of the setup, whilst the ion beam gradually mills into the overhanging structure. Once the milling has proceeded to the point where this tracking structure is flush with the porous GaN surface, the tomography proceeds through the ROI by capturing SE images of the porous surface as the ion beam mills to a further depths, and ends when the milling has removed the entire porous GaN layer, leaving NID GaN. 

Each frame of the tomographs in the plan-view tomography orientation is significantly larger. The thickness of the porous layer is fixed in all samples at 1\,\textmu m. For a theoretical three dimensional ROI with dimensions of 5\,\textmu m $\times$ 5\,\textmu m $\times$ 1\,\textmu m, imaged in conventional cross-sectional tomography, SE frames would need 5\,\textmu m of length and 1\,\textmu m of height, with the run proceeding until 5\,\textmu m of material had been milled along the milling direction/$\alpha$ axis. For the same size ROI imaged in plan-view tomography, each SE frame would have length of 5\,\textmu m and width of 5\,\textmu m, with the run instead proceeding until (at least) 1\,\textmu m of material had been milled along the milling direction. This aspect offers an additional advantage, as the dimensions of the deposited tracking structure are now significantly different. For this hypothetical ROI size, in cross-sectional tomography, the deposited tracking structure into which the five converging tracking markers would need to be 5\,\textmu m wide (along the $\beta$ axis) and 5\,\textmu m long (along the $\alpha$ axis), but in plan-view tomography, the same ROI would be captured ultimately by depositing a tracking structure that was 5\,\textmu m wide and only $\approx$ 2\,\textmu m long (including 1\,\textmu m of length for the entire porous layer and approximately 500\,nm of length extending in either direction onto the remaining shelf and into the NID GaN). This provides the additional advantage that the two converging tracking lines can be prepared with a significantly greater angle from the direction of the three parallel tracking marks. This then means that, for a given slice thickness, the tracking marks converge by a physically greater distance in the plan-view tomography setup than in the cross-sectional tomography setup, and the uncertainty in slice thickness decreases, such that the stabilisation period during the start of acquisition is faster. It is also notable that, since significantly fewer frames are captured, the longer duration of the setup stages for the plan-view tomography approach are compensated by a shorter overall acquisition time, such that the plan-view approach is not a prohibitively longer experiment overall.

Tomographs were captured in cross-sectional and plan-view orientations for the three samples in this work. Each presents a different set of challenges with which the tomography approaches can be evaluated. The manuscript will address each tomograph from both approaches captured on each of the three samples, separated into sections for each of the samples (sections \ref{8V carbonate}, \ref{4V carbonate} and \ref{8V oxalic} for Sample 1, 2 and 3, respectively). Beforehand, however, the metrics by which the accuracy and usefulness of a given tomograph must be discussed, in section \ref{evaluation}.

\section{Approaches to the Evaluation of Tomography Datasets}
\label{evaluation}

\begin{figure}
    \centering
    \includegraphics[width=0.5\linewidth]{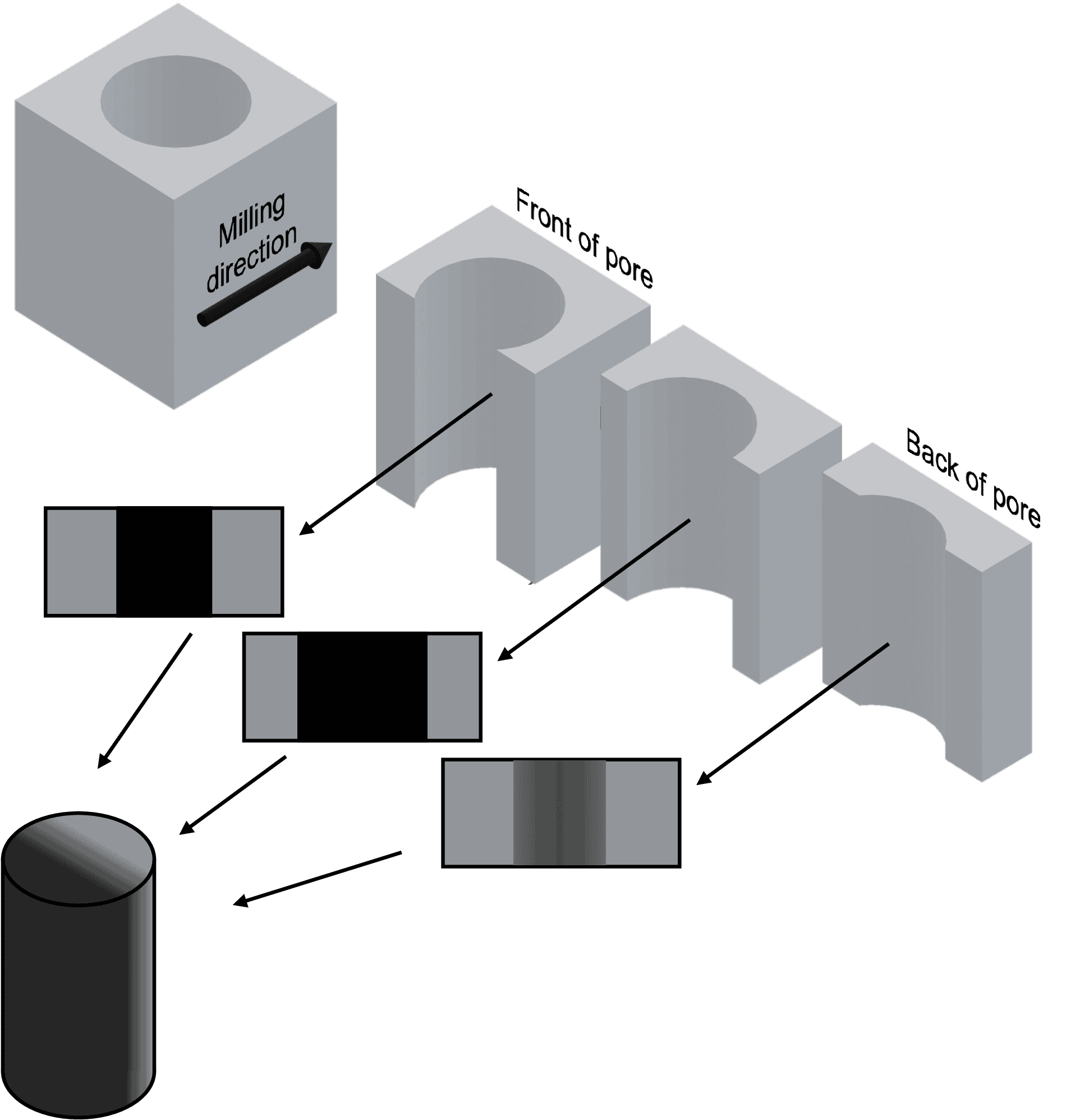}
    \caption{Schematic illustration of how the pore back effect leads to feature anisotropy and high intensity voxels. Images captured further along the milling direction have a shorter distance to the back wall of the pore, and thus register a greater intensity of detected electrons}
    \label{fig9}
\end{figure}

The two tomography approaches outlined in this work, and indeed all characterisation methodologies for any materials, can never be said to offer an absolute ground truth by providing entirely precise information as to microstructure. Most characterisation techniques are evaluated based on signal-to-noise ratio and feature resolution (which, in SE-based methods, is itself restricted by the hard limit of the interaction volume for the relevant beam condition and material). These issues certainly play a role in the absolute resolving power of FIB-SEM tomography, and are relevant aspects that will be discussed in the manuscript. However, the issues being addressed here are fairly unique - the pore back effect and voxel anisotropy present issues that are not necessarily well described or parametrised in literature. Indeed there are no clear methods for evaluating the pore back effect quantitatively, so this section will outline and justify some of the methods used in this work to appraise the fidelity of the tomographs in the two perspectives.

The first and most intuitive method for appraising a tomograph is to examine the reconstructed images that it produces from alternative perspectives to the imaging plane. Once aligned and registered, the voxels comprising the full three-dimensional tomograph may be collated to construct a `virtual image' from any specific two-dimensional plane that intersects the tomograph. These images are `virtual' insofar as they were not directly captured or imaged, \textit{per se}, but they are the result of compiling the voxels which contain structural information from positions in the plane to observe the same dataset from any arbitrary perspective. We recently demonstrated the value of this aspect of tomographic datasets by reconstructing plan-view images of the structure of pores in the porous layers of porous GaN DBRs, and used the resulting images to construct an entirely new framework for understanding the etching of these devices \cite{THORNLEY2026121957}. Thus, reconstructing virtual images from tomographs is an easy way to appraise whether the features are well resolved and distinct. When performing tomography in the conventional cross-sectional setup, the `input' frames (being the real as-taken SE images that are compiled together) are cross-sections of the porous layer, so reconstructed images from the cross-sectional datasets can be made that are from the plan-view (i.e. parallel to the porous GaN surface) perspective. Likewise, when performing tomography in the plan-view methodology, the input frames now are plan-view (i.e. from the same perspective as the virtual frames extracted from cross-sectional tomography) and the virtual images are cross-sectional (i.e. from the same perspective as the real input frames in cross-sectional tomography). This symmetry means that virtual reconstructed images from the tomographs in one orientation may be readily compared to the real input frames from the other orientation (since, in an ideal experiment, the two would be equivalent), but also compared to other imaging methods. For the cross-sectional tomographs, the reconstructed plan-view images at the surface can be compared to atomic force microscopy scans of the surface of the same samples. For the plan-view tomographs, the reconstructed cross-sectional images can be compared to cross-sectional micrographs of the same samples in SEM, not only using the input frames from the cross-sectional tomography (prepared by FIB milling) but also using images captured on cleaved cross-sections. These cleaved surfaces have some issues, such as non-planar structure and the possibility of non-representative porosity (since fracture surfaces during cleaving are expected to follow the weakest, and therefore most porous direction through the porous layer, and thus present pore morphology at non-arbitrary positions), but since the FIB-prepared surfaces may present some ion beam damage, the inclusion of both offers more certainty.

The morphological conformity of the reconstructed images to equivalent features imaged in alternative methods offers some general insight into the quality of the tomographic dataset. Extracting quantitative information from this approach, however, is not straightforward due to the lack of a perfect reference to which these reconstructed images may be quantitatively compared. The pore back effect will contribute significantly to the extent to which reconstructed images match with the comparator images, but there is no clear method for evaluating the extent of the pore back effect numerically in literature. This manuscript provides two methods for doing so. The first is based on the intensity distribution for the voxels in the porous layer, and the second on the uniformity of resolved features.

\begin{figure}
    \centering
    \includegraphics[width=0.9\linewidth]{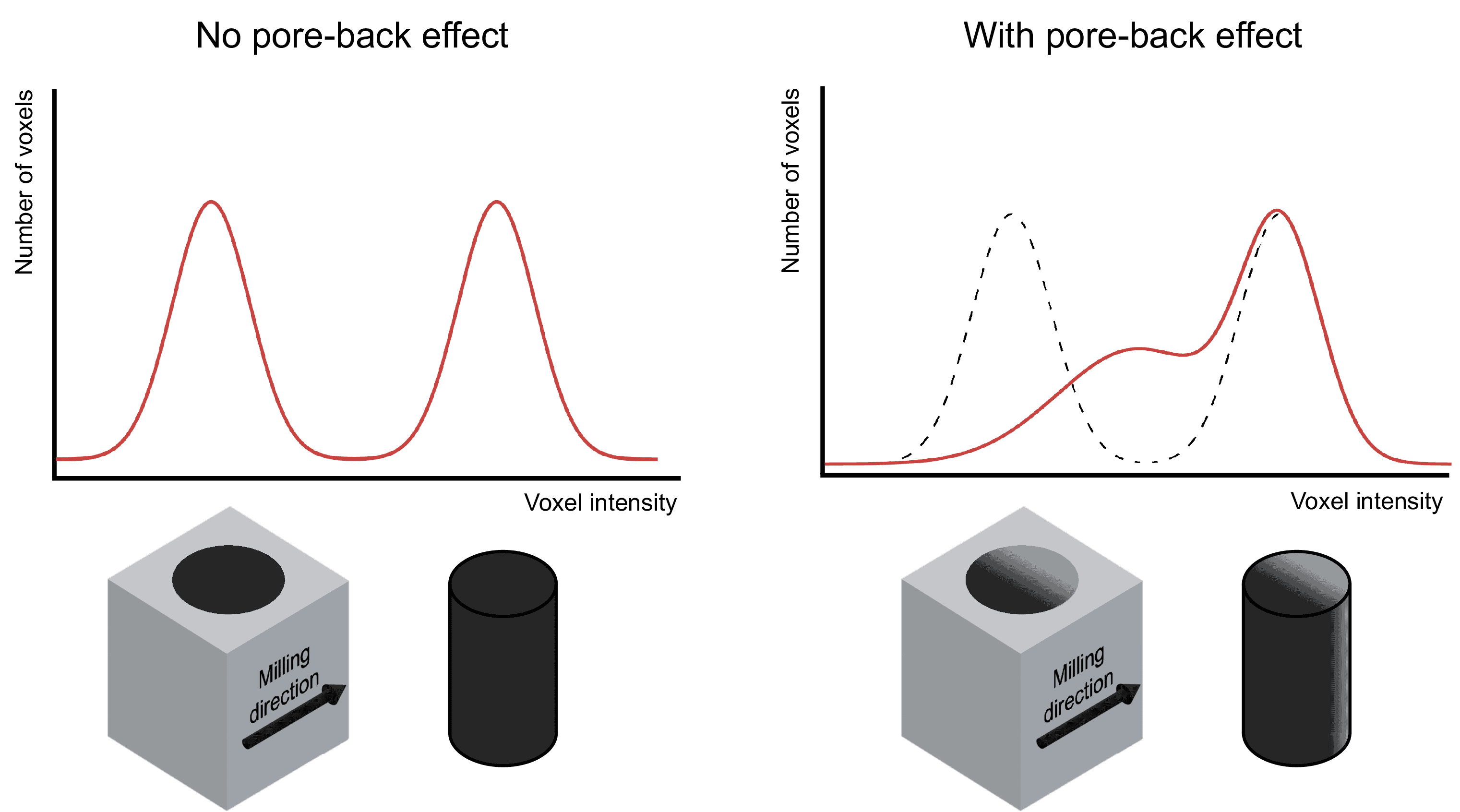}
    \caption{Schematic illustration of how the pore back effect distorts intensity distributions of tomographs. Low-intensity voxels are far less likely to manifest, such that the voxels associated with the pore class smear out into higher intensities and the overall distribution loses the separate intensity peaks}
    \label{fig10}
\end{figure}

A FIB tomography experiment performed on a binary composite material consisting of two solids would have a simple intensity distribution. Intensity, in this context, refers essentially to the brightness of the pixel on images, with a value between 0 (black) and 256 (white) assigned for every pixel of the input frames and, ultimately, for every voxel in the tomograph. Assuming that the two materials have significantly different intensity for the utilised imaging condition, it would be plausible to expect that, aside from noise and charging effects, all of the frames that were captured in the experiment would essentially be binary images, with one value of intensity in all spatial regions where one material was imaged and a different value for all areas where the other material was imaged. The total distribution of voxel intensity for the final result would, therefore, consist of two peaks, broadened out into Gaussian functions by the noise and uncertainties associated with the interaction volume of the imaging beam. Plotting this distribution as a histogram, one would expect there to be two peaks in intensity which entirely contained all of the voxels corresponding to each of the two materials.

Porous GaN is essentially a binary composite, consisting of GaN matrix and void. Were all of the pores in the material filled with a material with sufficient intensity contrast, the above would be true. However, void, unlike any other material in a composite, is completely transparent and allows features on the back walls to be imaged, which is the cause of the pore back effect. Hence, when imaging the void regions of the composite, the intensity is not consistent but depends on the distance between the imaging plane and the material at the back of the pore. This effect leads to a smearing out of the ideal intensity distribution along the milling direction, where, because the pores have a continuously increasing intensity when the back walls of the pores are moved successively closer to the imaging plane (as milling proceeds), the sharp peak of low intensity is instead spread out into higher intensities. It is thus appropriate to examine intensity distributions for the tomographs here - given that the same samples have been imaged with the same electron beam/detector configuration, the plan-view and cross-sectional orientations can be compared by comparing the intensity distributions. The more the distribution resembles the simple two-peak ideal case, the less the effect of the pore back has been. Likewise, the more that intensity can be seen to form a continuous distribution with no clear peak in the low intensity regime, the greater the effect of the pore back effect must have been. 

The other method used to appraise the extent of the pore back effect in this manuscript is to use the anisotropy of the features resolved with respect to the milling direction. Considering once again an ideal binary composite of solid materials, the pore back effect would not have any effect on isotropy of features. For a cylindrical inclusion of one material in the matrix, the tomograph would, not accounting for voxel dimensions, result in an isotropic feature. This is not true for a cylindrical pore in a porous material. When the pore is first imaged, and first reaches the imaging plane, the distance to the back wall is large and so the pore back effect is insignificant, and the intensity values are lower. Once the milling proceeds through the centre of the pore, however, symmetry is no longer maintained as the frames towards the back of the feature will be more affected by the pore back effect than those at the front. Thus, for a given distance from the centre of the cylinder in either direction along the $\alpha$ axis, it is not possible for the same contrast to be captured despite the pore itself having the same intersection with the imaging plane. For a binary composite, these frames would be the same, but for a porous material, the image captured closer to the back wall of the pore would have a greater intensity and more shine-through in the porous region than the one captured closer to the front wall.

The overall effect of this is to introduce anisotropy to the tomograph that does not necessarily exist in the structure of the pores. Isotropic features such as cylinders will be compressed along the back wall and appear distorted. By extracting line profiles in intensity, the symmetry of features can be compared. A line profile along the milling direction/$\alpha$ axis will show asymmetry that does not exist in line profiles perpendicular to the milling direction (i.e. along the $\beta$ axis) when the pore back effect is severe. We note that whilst pores in some of these samples do not necessarily create an isotropic shape when they intersect with the imaging plane, by examining the extent of distortion along the back walls of pores using these extracted line profiles, the extent of the effect may be numerically understood. 

\section{Results and discussion}
\label{R+D}

\subsection{Sample 1}
\label{8V carbonate}

\begin{figure}
    \centering
    \includegraphics[width=1\linewidth]{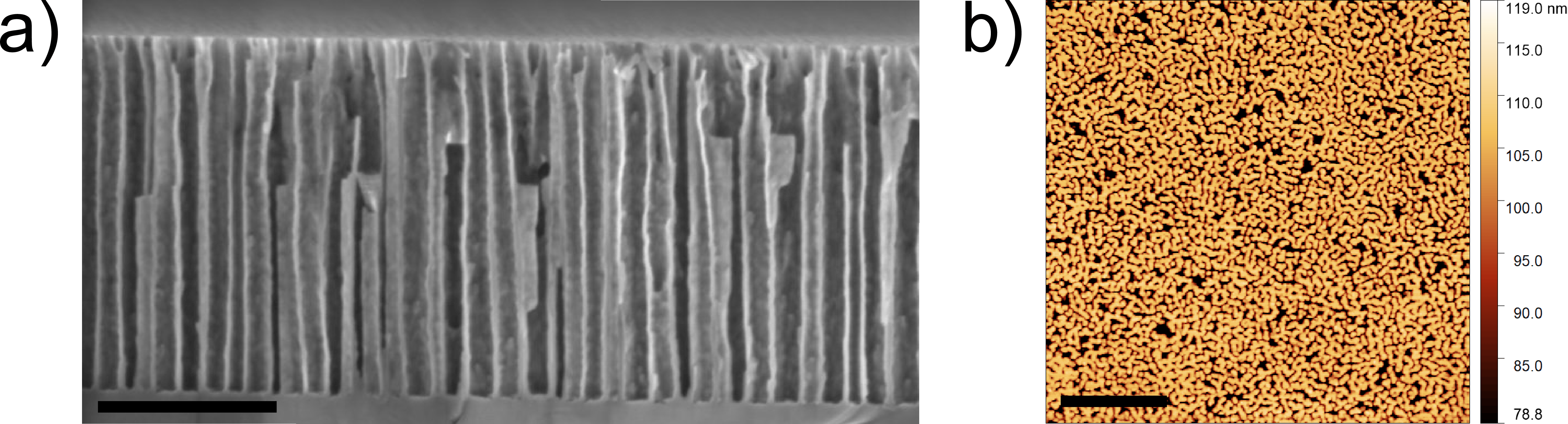}
    \caption{Reference images for the structure of pores in Sample 1, showing a) cross-sectional SEM of a cleaved edge of the porous layer and b) AFM of the surface of the porous layer. Scale bars represent 500\,nm.}
    \label{sample1Ref}
\end{figure}

Sample 1 was etched in a 0.1\,M aqueous solution of Na$_2$CO$_3$ at a potential of 8\,V in a 3-electrode setup. The structure may be understood as consisting of wide and straight columnar pores. Cross-sectional SEM on a cleaved edge of the sample and an AFM scan are both given in Figure \ref{sample1Ref}, where the features are illustrated. AFM scans take place on the top surface of the porous layer. Figure \ref{sample1Ref}a gives an SE image of a cleaved cross-section and Figure \ref{sample1Ref}b gives the AFM scan of the surface. It is of note that not every pore in the sample remains entirely columnar, and that a small number of them bifurcate into two smaller pores as etching proceeds downwards. It is also worth noting that the surface pore structure is entirely different to the structure in the bulk. The surface is dominated by a large number of fine pores, only a minority of which persist downwards into the main body of the layer. 

Tomographs captured on this sample are provided in their entireties in the supplementary material, in the form of videos. `Supplementary Video 1 - Sample 1' provides both tomographs on Sample 1, in the form of four distinct animations. The first animation is a fly-through of the input frames of the cross-sectional tomograph; the second is then the reconstructed plan-view virtual images from the cross-sectional tomograph, animated to move from the surface of the GaN downwards through the layer. The third is then the input frames from the plan-view tomograph (which are from an equivalent perspective to the second, but are as-taken, rather than reconstructed virtual images) animated to move from the surface of the porous GaN layer downwards through the layer. Finally, the fourth animation is the reconstructed virtual cross-sectional images from the plan-view tomograph animated to sweep from one (arbitrary) edge of the ROI to the other. For both tomographs, the image pixel size was 1\,nm, and the average slice thickness differed only slightly, being 5.7\,nm for the cross-sectional tomograph and 5.9\,nm for the plan-view tomograph.

\begin{figure}
    \centering
    \includegraphics[width=1\linewidth]{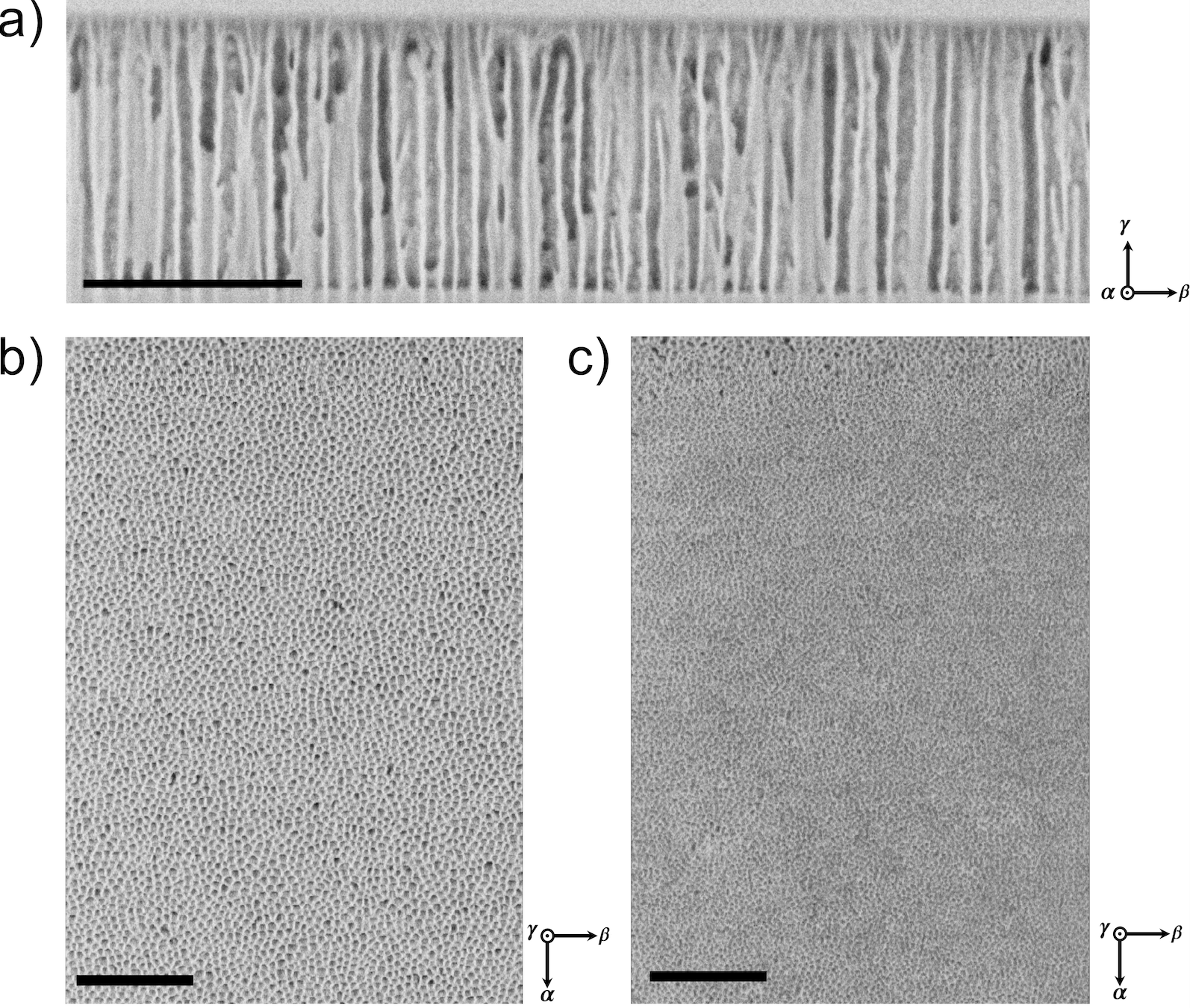}
    \caption{Summarised results from cross-sectional tomography of Sample 1, showing a) an example of an input frame, b) a reconstructed plan-view virtual image positioned at a depth along the $\alpha$ axis of approximately 500\,nm and c) a reconstructed plan-view virtual image positioned at the surface of the porous layer. All scale bars represent 1 \textmu m on the bottom-left of the image and all images have related axes on the bottom-right side.}
    \label{8VC_XStomo}
\end{figure}

The results of the cross-sectional tomography are summarised in Figure \ref{8VC_XStomo}. Figure \ref{8VC_XStomo}a shows a typical input frame from the dataset from the centre of the ROI, whilst Figure \ref{8VC_XStomo}b then shows a typical reconstructed plan-view virtual image of the centre of the porous layer. Finally, Figure \ref{8VC_XStomo}c shows another reconstructed plan-view virtual image, instead positioned to show the surface of the porous layer as reconstructed from the tomograph. 

These images may be compared to appraise the fidelity of the tomograph. It is first worth comparing Figure \ref{sample1Ref}a and Figure \ref{8VC_XStomo}a, which are both as-taken images of the same sample from the same perspective. Figure \ref{sample1Ref}a, taken in normal SE imaging of a cleaved cross-section shows significantly less noise than the example tomography frame in Figure \ref{8VC_XStomo}a. This is a result of the longer dwell time and finer pixel size used in Figure \ref{sample1Ref}a. It becomes impractical to capture tomographs at such long imaging durations, whereas Figure \ref{sample1Ref}a, intended to show the structure, was captured to maximise the fidelity. The other striking difference is that the tomography input frame in Figure \ref{8VC_XStomo}a has pores which show more variation in intensity than the pores in Figure \ref{sample1Ref}a. This is likely to be a demonstration of the pore back effect; the tomography frame will be captured along an arbitrary milled plane, whereas the image on the cleaved surface will not, since cleaving will naturally take place along the weakest (i.e. most porous) direction. Hence, the cleaving surface will likely have a trajectory that mostly intersects the centres of the pores on the surface, whilst the tomography frame will intersect the participating pores at arbitrary distances through their widths. Hence, Figure \ref{8VC_XStomo}a shows significantly more of this variation in the intensity of the pores than Figure \ref{sample1Ref}a. It may be suggested that the preference of cleaving to create fracture surfaces along highly porous planes through the layer also selects for larger pores, such that the features imaged on cleaved samples are generally larger than average, though this does not appear to be the case as Figure \ref{sample1Ref}a and Figure \ref{8VC_XStomo}a generally demonstrate high conformity. 

Figure \ref{8VC_XStomo}b shows the results of tomographic reconstruction from cross-sectional frames to appraise pores from the plan-view perspective, at the centre of the ROI (i.e. at a depth of around 500\,nm below the surface of the porous GaN). The scanning direction was along the negative $\alpha$ axis, such that the frames captured earlier on contribute to the tomograph lower down on this reconstructed image. The image shows a reasonable conformity to the expected pore morphology; dark patches represent columnar pores being intersected in the plan-view imaging plane. There is a reasonable diversity of pore size and shape but the structure is generally uniform, which is consistent with data from other imaging modes. The pore back effect and resultant anisotropy are apparent here; however, examination of the pores throughout Figure \ref{8VC_XStomo}b shows that a great many of them exhibit anisotropy along the (negative) $\alpha$ direction, where the top of the pore is darker than the bottom. This is a clear example of the effect as described in section \ref{evaluation} and will be discussed in the context of the line profiles extracted from this image later on. Finally, Figure \ref{8VC_XStomo}c shows another reconstructed plan-view virtual image, in this case positioned higher to show the surface of the porous layer. There is a fine network of porous features with low contrast from the surrounding GaN and poor feature resolution. On closer inspection, this shows some extent of conformity to Figure \ref{sample1Ref}b, but given the small feature size and high extent of noise, it cannot be claimed that this image offers a high-quality reconstruction of the pore structure at this position. There is a small region at the top of the image where the pores are larger, though this is an artefact of poor alignment rather than a failure of tomography. It is also notable that the surface pores do not exhibit the same feature anisotropy as the pores in Figure \ref{8VC_XStomo}b in the bulk of the layer, though this is likely due to the fact that the pores at the surface are so small that very few frames were compiled together during the reconstruction of any one of them. 

\begin{figure}
    \centering
    \includegraphics[width=1\linewidth]{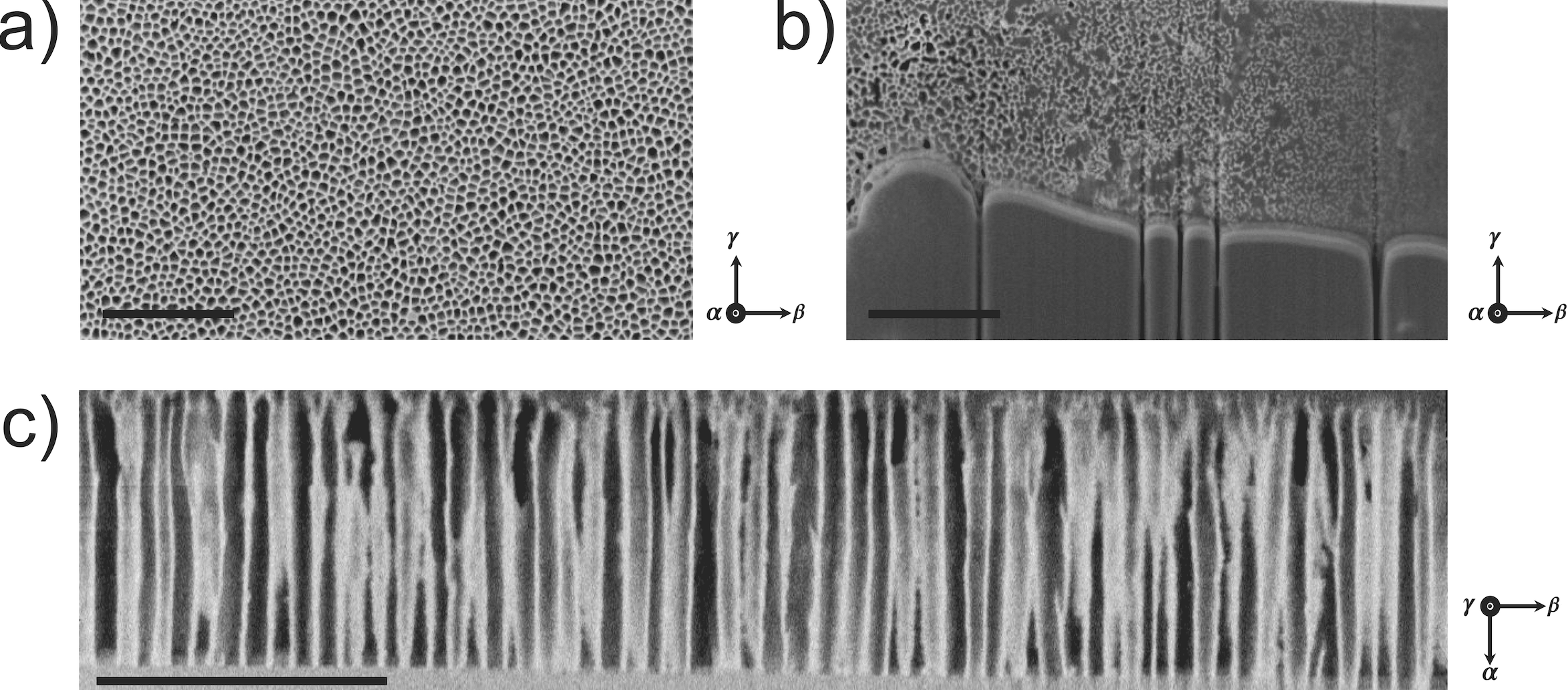}
    \caption{Summarised results from plan-view tomography of Sample 1, showing a) an example of an input frame, b) an early input frame featuring some of the surface of the porous layer, c) a reconstructed cross-sectional virtual image positioned at the centre of the porous layer. All scale bars represent 1\,\textmu m on the bottom-left of the image and all images have related axes on the bottom-right side.}
    \label{8VC_Vtomo}
\end{figure}

The results of the plan-view tomography on Sample 1 are summarised in Figure \ref{8VC_Vtomo}, somewhat analogously to Figure \ref{8VC_XStomo}. Figure \ref{8VC_Vtomo}a shows an example of an input frame from the plan-view perspective, taken from the centre of the porous layer, showing the columnar pores as tessellating polygons with strong contrast between the voids and the GaN. Figure \ref{8VC_Vtomo}b shows another input frame, in this case positioned to show a region of the surface of the porous layer. This image has several other features due to some external effects during capture - the frame does not feature the entire surface due to the fact that the sample was not mounted at exactly 90$^\circ$ to the stage during tomography, meaning that the porous layer emerges through the top of the image at an earlier point than the at the bottom of the image. We note that the full surface could still be imaged by reconstructing an image from that tilted angle but that the image included is an as-taken one for consistency. The layer is also more porous on the left-hand side than the right, again likely due to imperfect mounting. Finally, some curtaining issues are observed as five vertical lines which emerge due to imperfect filling of the tracking marks with carbon, creating voids in the overlaying tracking structure at which curtains nucleate. This is unfortunate but is not a feature of plan-view tomography specifically and can be somewhat mitigated by applying destriping algorithms during image processing. Finally, Figure \ref{8VC_Vtomo}c shows a reconstructed cross-sectional image from the centre of the region of interest (arbitrarily). There is a small tilt from the left to the right, again due to the sample being mounted at an angle, but otherwise this image is entirely as expected.

The two approaches to tomography may be compared to one another by assessing the extent to which the outputted reconstructed images offer a meaningful reflection of the true pore morphology. For cross-sectional tomography, the reconstructed plan-view image in Figure \ref{8VC_XStomo}b (at the centre of the layer) may be compared to the as-taken image now captured in Figure \ref{8VC_Vtomo}a. We can also compare the reconstruction of the surface in Figure \ref{8VC_XStomo}c to the AFM scan of the surface in Figure \ref{sample1Ref}b. For plan-view tomography, the reconstructed cross-sectional image in Figure \ref{8VC_Vtomo}c may be compared to the as-taken cross-sectional images in Figure \ref{sample1Ref}a and Figure \ref{8VC_XStomo}a.

The cross-sectional tomograph does a reasonable job of reflecting the true structure of pores. Comparing the reconstruction in Figure \ref{8VC_XStomo}b and the captured image in Figure \ref{8VC_Vtomo}a, the general structure is well represented with pores showing approximately the same scale and size monodispersity. The biggest issue is the feature anisotropy - there is a large range of intensity across the pore in figure \ref{8VC_XStomo}b, making it difficult to discern the exact positions of pore walls and impossible to infer any texture or roughness on the walls of the pores, as well as to segment the pores for any quantitative analysis. Likewise, the cross-sectional tomograph has very limited capability to reconstruct the surface of the porous layer; comparing Figure \ref{8VC_XStomo}c and Figure \ref{sample1Ref}b shows that there is a weak extent of feature capture, although the outline of a similar pore morphology can be partially observed. The plan-view tomograph, on the other hand, does a very good job of reconstructing the cross-sectional perspective. Figure \ref{8VC_Vtomo}c shows very good conformity to Figure \ref{sample1Ref}a, showing the roughness of pore walls with good contrast. Indeed, the pores appear considerably darker in the reconstructed virtual image (except where the imaging plane partially intersects with pore walls) than in the as-taken image from the cross-sectional tomograph in Figure \ref{8VC_XStomo}a. Despite the fact that both tomographs were captured with the same imaging conditions, the reconstruction in Figure \ref{8VC_Vtomo}c shows the pores with significantly more clarity than the image in Figure \ref{8VC_XStomo}a. This is because the reconstruction, which does not suffer significantly from the pore back effect, has darker voxels in the pores than can be accessed by cross-sectional imaging. This drastic difference from a simple rotation of the imaging modality demonstrates the severe impact that the pore back effect can have on such experiments. 

\begin{figure}
    \centering
    \includegraphics[width=0.7\linewidth]{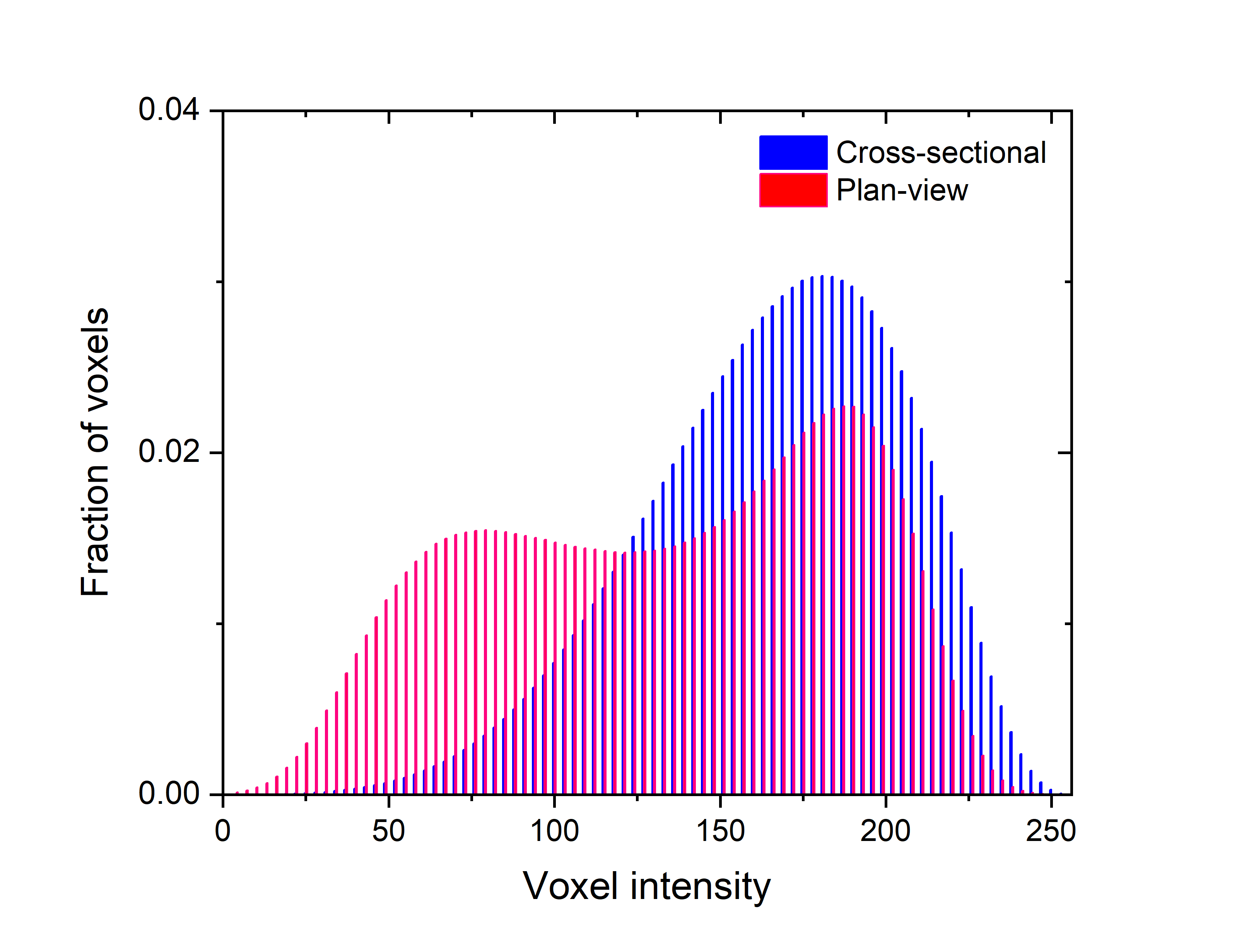}
    \caption{Histograms of distributions of voxel intensity values to compare cross-sectional (blue) and plan-view (red) tomography of Sample 1, showing the value of intensity plotted against the fraction of total voxels in the two tomographs with that intensity}
    \label{sample1Histograms}
\end{figure}

This comparison of overall intensity may be facilitated by comparing the overall distributions of voxel intensity. Figure \ref{sample1Histograms} compares the intensity distributions for the two tomographs on Sample 1. These distributions are also available plotted on a logarithmic y-axis in the supplementary material (Supplementary Figure 1). The perfect behaviour would be to have two discernible peaks in intensity, one representing all of the voxels corresponding to GaN and another representing those corresponding to void, as discussed in section \ref{evaluation}. Any instance of the pore back effect will have the effect of smearing out this low intensity peak to higher intensity values. 

The two distributions in Figure \ref{sample1Histograms} have a clearly different form to one another. The intensity histogram for the cross-sectional tomograph shows a continuous single peak, with very few voxels in the lower intensity range. The pore back effect has had such an impact on the dataset that there is no clear peak corresponding to the void at all, since the intensity in pores must have been sufficiently high to smear the distribution into a single peak. By contrast, the plan-view tomograph has an intensity distribution with a clear peak in the low intensity range, and significantly more voxels in the low intensity range in general. These two peaks are not completely separate, as would be the ideal case, but it is still evident that the GaN and void peaks are distinguishable due to the significant mitigation of the pore back effect. This shows fairly straightforwardly that the plan-view approach offers a significant advantage for mitigating the pore back effect.

\begin{figure}
    \centering
    \includegraphics[width=0.825\linewidth]{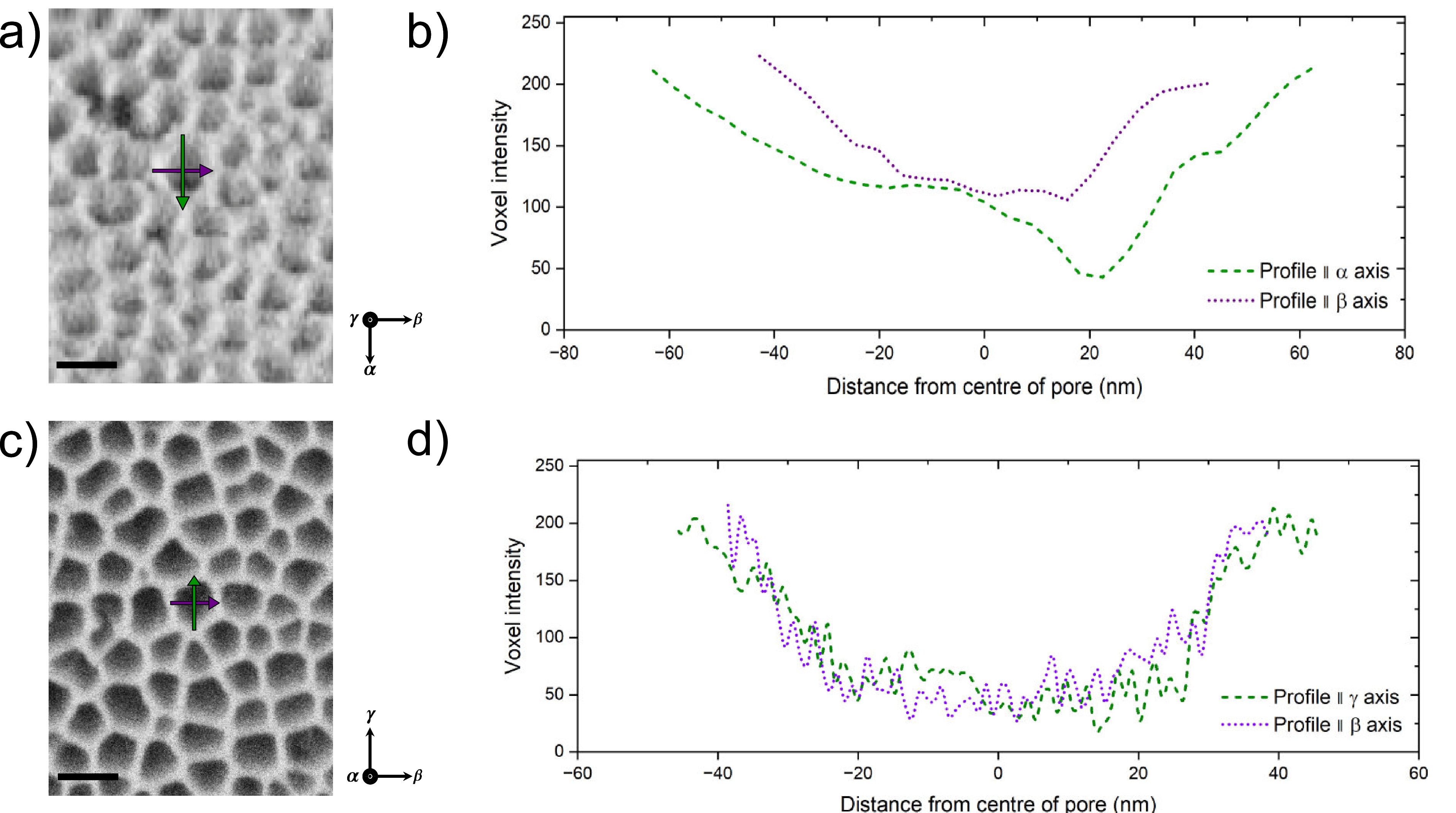}
    \caption{An illustration of feature anisotropy of Sample 1 for the cross-sectional tomograph, showing a) an extracted region of the reconstructed plan-view image from Figure \ref{8VC_XStomo}b highlighted with the position and direction of line profiles and b) the extracted line profiles, compared to the equivalent as-taken images from the plan-view tomograph shown as c) an extracted region of the as-taken plan-view image from Figure \ref{8VC_Vtomo}b highlighted with the position and direction of line profiles and d) the extracted line profiles. Scale bar represents 100\,nm}
    \label{Lineprofilessample1}
\end{figure}

Feature anisotropy due to the pore back effect can be appraised using line profiles extracted from reconstructed images. An example of this is given in Figure \ref{Lineprofilessample1}, where a pore in the reconstructed plan-view image extracted from the cross-sectional tomograph on Sample 1 (which has been previously shown in Figure \ref{8VC_XStomo}b) has been processed to extract line profiles of voxel intensity. Figure \ref{Lineprofilessample1}a shows the pore in question highlighted with two arrows indicating the directions and positions of two line profiles, both passing through the (approximate) centre and given in Figure \ref{Lineprofilessample1}b. These line profiles are presented side-by-side with the equivalent line profiles from the plan-view tomograph relative to the sample dimensions (i.e. the line profiles are extracted from as-taken images from plan-view tomographs, since they are the equivalent pore geometry). This can be used as a demonstration of the pore back effect and shows unambiguously the effects that the pore back effect can have. Thus, Figure \ref{Lineprofilessample1}c is extracted from a small region of Figure \ref{8VC_Vtomo}b, and the line profiles are now parallel to the $\gamma$ and $\beta$ axes, since this is not a reconstructed image. It is readily apparent that the line profile taken parallel to the $\alpha$ axis is significantly more anisotropic, as the profile is lower in symmetry than the one taken parallel to the $\beta$ axis. The lowest intensity value of this line profile is not at the centre of the feature, as would be expected, but instead around 20\,nm further along the (positive) $\alpha$ axis, corresponding to an earlier frame where the distance to the pore back is large. Since there is generally symmetry for the line profile extracted parallel to the $\beta$ axis, and there is no reason to believe that the sample demonstrates any fundamental structural anisotropy along this direction, this can be attributed to the pore back effect. It is also notable that this asymmetry is not present in the line profiles of the as-taken image from the plan-view tomograph in Figure \ref{Lineprofilessample1}d, where the two are approximately the same. It is notable that, since the instrument has an angle between the ion and electron beams of 54$^\circ$, there could be expected to be some shading on the bottom of the pores due to the angle of imaging (that would not be there if the electron beam was installed at 90$^\circ$ to the ion beam, and was thus perpendicular to the milled surface during imaging). However, this does not appear to be a major hindrance to features of this size. The line profiles in Figure \ref{Lineprofilessample1}d also demonstrate clearer definition (i.e. a shorter distance over which the intensity drops) of pore walls than those in Figure \ref{Lineprofilessample1}b, showing that these images are easier to use for segmentation of pore features than the reconstructed images from the cross-sectional tomographs.

We note that this is not necessarily a metric by which the plan-view and cross-sectional tomographs can be fairly compared. To compare these, line profiles would need to be extracted between the front and back walls of pores in the plan-view imaging modality. Since the imaging begins at the pore surface, which is coated with carbon, there is no front wall of the pore with GaN. Likewise, for this sample, the back wall is where the pores terminate at the bottom of the porous layer, which is up to 1 \textmu m away from the pore surface, meaning line profiles would be very long. Thus, these line profiles are introduced as a metric for evaluating the extent to which the pore back effect exists in general, but only to be applied to suitable morphologies, for which the plan-view tomograph of Sample 1 is not suitable. It may be of value, for example, to conduct a study comparing tomography datasets with different electron beam/detector configurations and using this method to appraise the extent of the pore back effect. 

On this sample, it is fairly clear that the plan-view tomography offers considerable advantages for feature resolution. The following two sections move to other pore morphologies which both, for separate reasons, push the limits of the resolution of the tomography experiments, to appraise whether the advantages of this approach carry over to very different pore morphologies. 

\subsection{Sample 2}
\label{4V carbonate}

\begin{figure}
    \centering
    \includegraphics[width=1\linewidth]{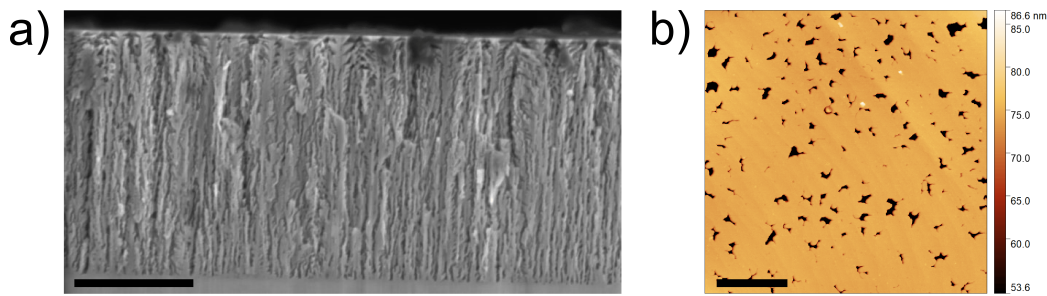}
    \caption{Reference images for the structure of pores in Sample 2, showing a) cross-sectional SEM of a cleaved edge of the porous layer and b) AFM of the surface of the porous layer. Scale bars represent 500\,nm.}
    \label{sample2Ref}
\end{figure}

Sample 2 was etched in a 0.1\,M aqueous solution of Na$_2$CO$_3$ at a potential of 4\,V in a 3-electrode setup. The structure may be understood as being extremely narrow columnar pores through the layer. They have a significantly different structure at the surface, where the surface consists of small polygonal pits that are widely spaced. In the first 50\,nm of the porous layer, moving downwards, these polygonal pores expand outwards into star-shaped features which then continue as a network of fine vertical pores with very fine branches emanating a short distance around them. Again, Figure \ref{sample2Ref}a gives an SE image of a cleaved cross-section and Figure \ref{sample2Ref}b gives the AFM scan of the surface. 

Again, both tomographs captured on this sample are provided in their entireties in the supplementary material, as `Supplementary Video 2 - Sample 2', which shows four different perspectives analogously to Supplementary Video 1, as described in section \ref{8V carbonate}. For both tomographs, the image pixel size was 1\,nm, and the average slice thickness was 5.8\,nm for the cross-sectional tomograph and 5.7\,nm for the plan-view tomograph.

\begin{figure}
    \centering
    \includegraphics[width=1\linewidth]{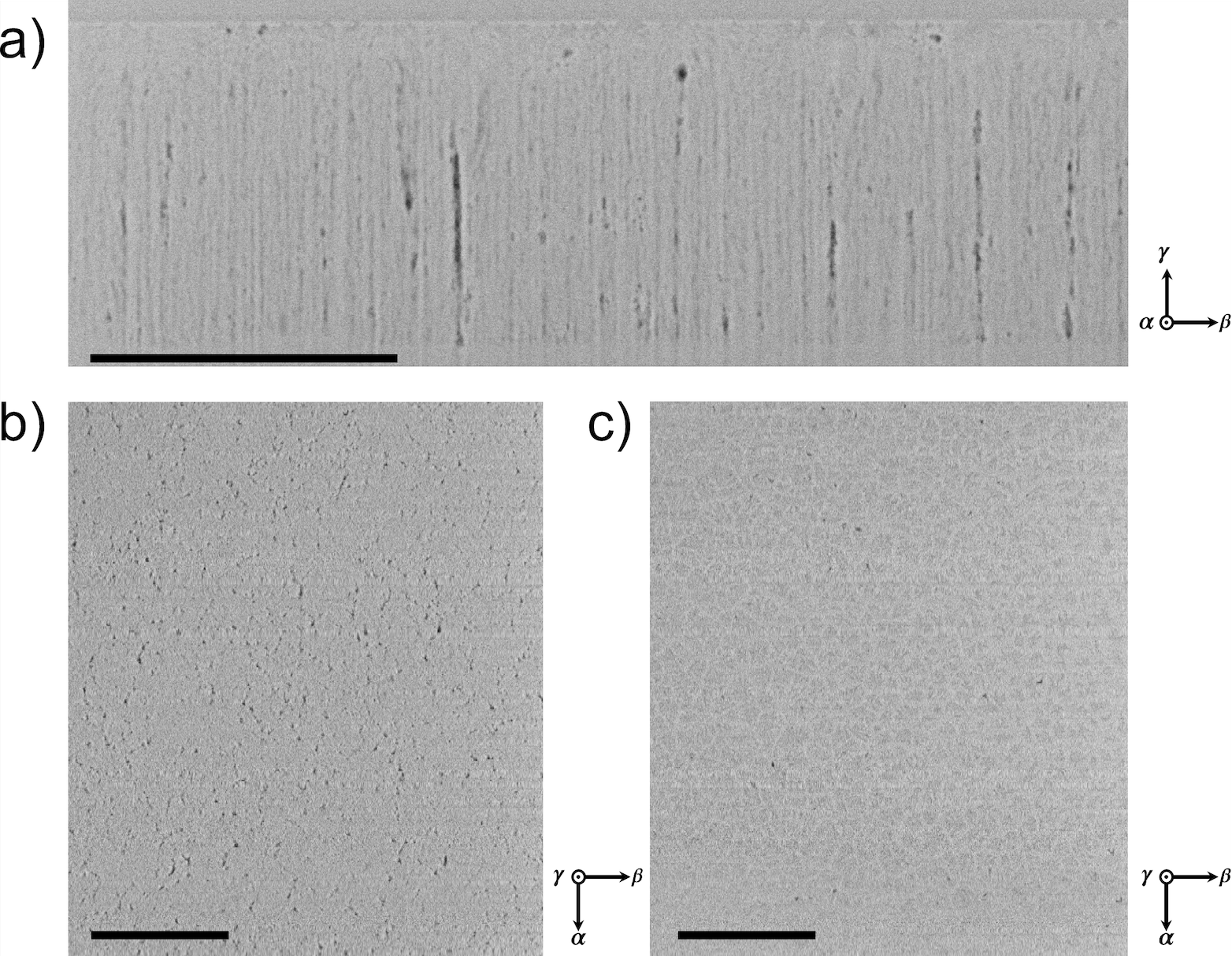}
    \caption{Summarised results from cross-sectional tomography of Sample 2, showing a) an example of an input frame, b) a reconstructed plan-view virtual image positioned at a depth along the $\alpha$ axis of approximately 500\,nm and c) a reconstructed plan-view virtual image positioned at the surface of the porous layer. All scale bars represent 1 \textmu m on the bottom-left of the image and all images have related axes on the bottom-right side.}
    \label{4VC_XStomo}
\end{figure}

Results of cross-sectional tomography on Sample 2 are summarised in Figure \ref{4VC_XStomo}, analogously to Figure \ref{8VC_XStomo}, with the input frame in Figure \ref{4VC_XStomo}a, the reconstructed plan-view virtual image at the centre of the porous layer in figure \ref{4VC_XStomo}b and the reconstructed plan-view virtual image of the porous layer surface in figure \ref{4VC_XStomo}c. 

The example cross-sectional frame in Figure \ref{4VC_XStomo}a shows some conformity to the cleaved cross-sectional SE image in Figure \ref{sample2Ref}a, but a lot of the finer detail is lost. Since this FIB-milled surface will be in an arbitrary direction, it is not a surprise that some of the pores have very little contrast to the GaN, since they will again lie at a variable distance from the surface, unlike the cleaved sample. For this sample, however, the pore diameters are so small that the depth over which they are likely to show high contrast is very little, hence why the features in this sample are so poorly discernible. There is some detail captured as to the texture of the pore walls, but not at the resolution in Figure \ref{sample2Ref}a. It is possible that amorphisation damage from the ion beam plays a role here, but a more systematic study using varying ion beam acceleration voltages would be needed to corroborate this claim. 

Figure \ref{4VC_XStomo}b shows a reconstructed plan-view image of the centre of the porous layer for the cross-sectional tomography. For this sample there is a fair diversity in the overall intensity of the features and in general, they are difficult to discern from the GaN. The likely reason for this is that the pores are so fine that they are only imaged across a small number of frames (2-5). Thus, for the pore to appear dark, it must not only have a large size to increase the distance to the back wall during imaging (and mitigate the pore back effect) but also to be milled favourably, such that the frames across which it is imaged happen to take place at milling depths where the distance is sufficiently large. In general this offers a rather restricted view of the pore morphology. As for the surface of the porous layer, reconstructed in Figure \ref{4VC_XStomo}c, it can be observed that there is some extent of morphological conformity between the reconstruction and the AFM scan in Figure \ref{sample2Ref}b, where there are some observable polygons, but there is such low contrast between these features and the GaN that this insight is limited. These features are also observed to fill space much more than the exact surface features observed in the AFM scan in Figure \ref{sample2Ref}b, suggesting that the features in the very shallow sub-surface cannot be adequately distinguished for this tomograph. The pore morphology does vary significantly at the surface over an extremely shallow depth range, so it may be unfair to compare the conformity between the surface reconstruction and AFM scan for this sample specifically, but in general, the reconstructing power of this dataset is extremely limited. 

\begin{figure}
    \centering
    \includegraphics[width=1\linewidth]{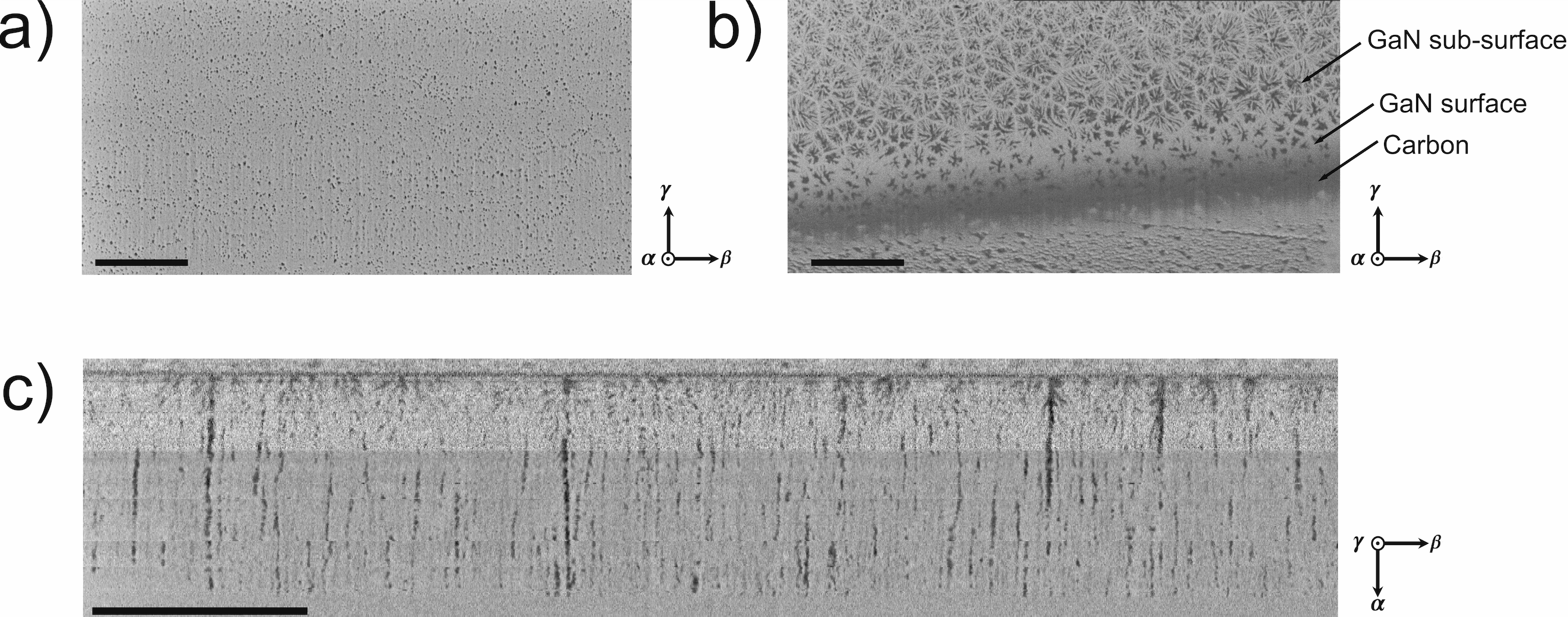}
    \caption{Summarised results from plan-view tomography of Sample 2, showing a) an example of an input frame, b) an early input frame featuring some of the surface of the porous layer, c) a reconstructed cross-sectional virtual image positioned at the centre of the porous layer. All scale bars represent 1 \textmu m on the bottom-left of the image and all images have related axes on the bottom-right side.}
    \label{4VC_Vtomo}
\end{figure}

The results of plan-view tomography on Sample 2 are summarised in Figure \ref{4VC_Vtomo}, again analogously to Figure \ref{8VC_Vtomo}, with the example input frame in Figure \ref{4VC_Vtomo}a, the example of the input frame near to the surface in \ref{4VC_Vtomo}b and the reconstructed cross-sectional image in Figure \ref{4VC_Vtomo}c. Examination of the input in Figure \ref{4VC_Vtomo}a shows the expected morphology - the pores now appear as very small dark dots which have a spacing significantly greater than their diameter. There is also a certain extent of curtaining, appearing as vertical lines, likely the result of the low porosity and fine pore features. Figure \ref{4VC_Vtomo}b again shows an example of an inputed frame close to the porous GaN surface, showing a similar sample tilting from the mounting. This time, the overhang angle is greater. The features at the top of the image are deeper into the porous layer. This image is actually very useful for communicating the structure near the surface; at the bottom of the image, the surface consists of the deposited carbon. Above the dark band is the surface of the porous layer, showing the polygonal pores that have the same morphology shown in the AFM scan of the surface in Figure \ref{sample2Ref}b. Above these on the image (and thus deeper into the layer), the polygonal pores are seen to expand outwards into branched cells that fill space with fine tipped branches, and these tips eventually form needle-like vertical pores with the extended morphology shown in Figure \ref{4VC_Vtomo}a.

Figure \ref{4VC_Vtomo}c shows a reconstructed virtual cross-sectional image from this plan-view tomograph, where the dark grey features are pores that intersect the (arbitrarily positioned in the centre) imaging plane. There is a horizontal band at the top, where the top of the porous layer appears brighter than the bottom, and some minor periodic disturbances on the same direction. From looking at the composite frames, this appears to be a software glitch, possibly reflecting an unprompted change in the brightness/contrast or some imaging condition due to an error. Since there is good continuity of all of the features, the data is presented raw, but these errors could of course be treated by image processing. Generally speaking, this reconstructed image does a reasonable job of reconstructing the cross-sectional perspective, though not as well as for Sample 1. Since the image plane is arbitrary, and the pores are of a far lower density, the number of pores on this image is significantly lower than on the cleaved surface imaged in Figure \ref{sample2Ref}, but the general form translates well. Some amount of texture on the pore walls can be resolved, showing the branching, though this is not true for some of the smaller features. The pore morphology at the top of the layer has also been well captured, due to the preferential imaging condition in this orientation. 

Thus, in comparing the two tomography approaches by the extent to which they can reconstruct mutual orthogonal perspectives, the plan-view approach seems to succeed better than the cross-sectional approach, though both suffer more to capture this extremely fine morphology. The improved contrast along the long axis of the pores contributes significantly to the improved resolution of pore morphology, both in the bulk and at the crucial surface region where pore morphology tends to evolve rapidly. With refinements to the slice thickness and a greater imaging time, it is plausible that the plan-view tomograph could be further improved, capturing the roughness of the pore walls to a greater extent and reducing noise elsewhere. For the cross-sectional tomograph, it is unlikely that any improvements could be made, since the bottleneck is mostly the pore back effect. 

\begin{figure}
    \centering
    \includegraphics[width=0.7\linewidth]{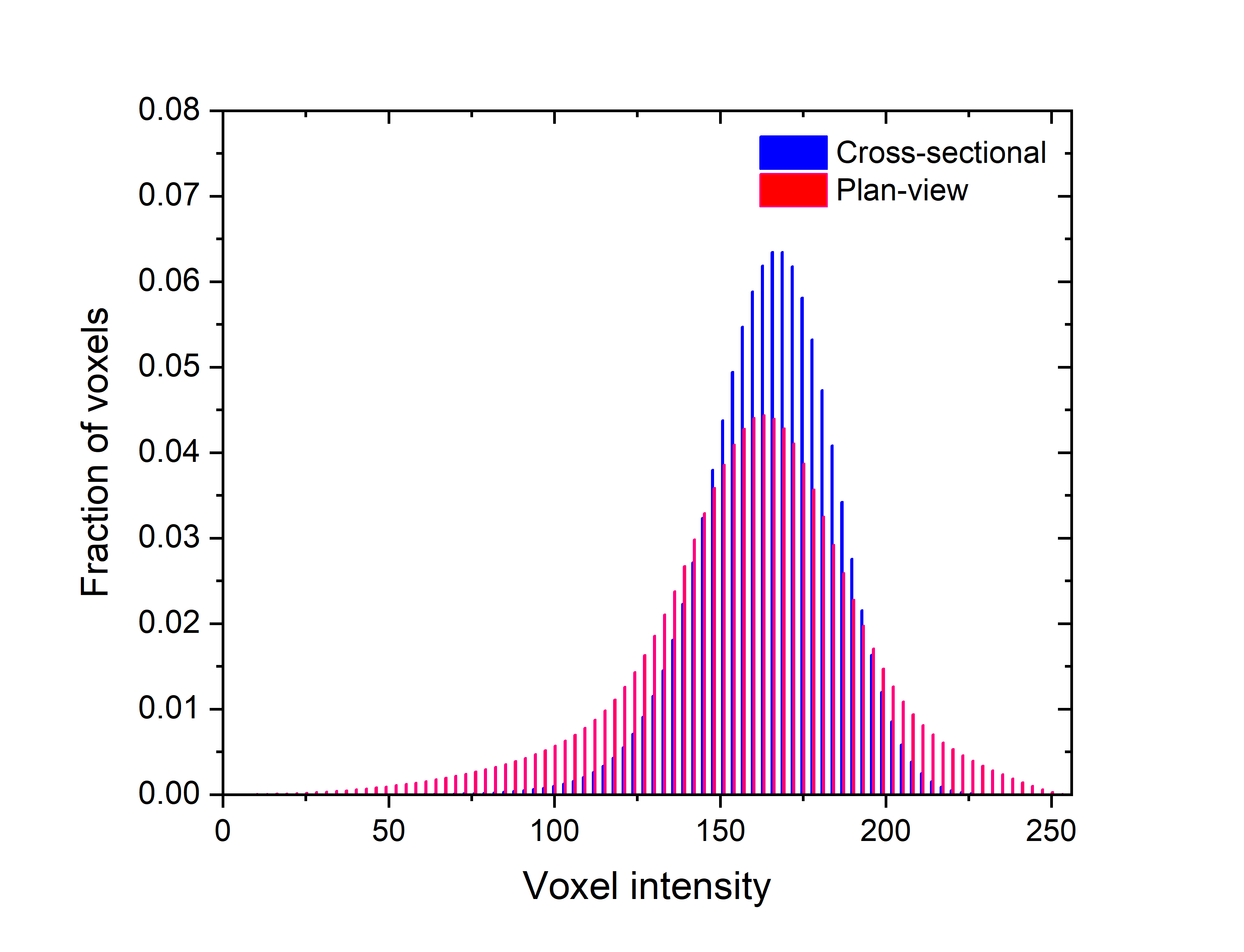}
    \caption{Histograms of distributions of voxel intensity values to compare cross-sectional (blue) and plan-view (red) tomography of Sample 2, showing the value of intensity plotted against the fraction of total voxels in the two tomographs with that intensity}
    \label{sample2histograms}
\end{figure}

Histograms of intensity distributions are given for these two approaches in Figure \ref{sample2histograms}. These both now show the same form, possessing one distinct peak with no evidence of the ideal two-peak behaviour. The plan-view approach, however, has broadened the peak out into the low intensity range, and generally there are more voxels with a lower intensity value in this distribution, reflective of the mitigation of the pore back effect. One of the major factors contributing to the plan-view tomograph moving away from the two-peak form is the low porosity, such that the actual volume fraction of voids being significantly lower reduces the size of the void peak significantly. Indeed, when these data are viewed with a logarithmic scale on the y-axis, a slight shoulder does appear for the plan-view data which is not present for the cross-sectional data. This is available in the supplementary material as Supplementary Figure 2. However, it is a stretch to say that this shoulder is significant proof to the effect of the formation of the second peak, though it certainly hints at that effect. Comparing these histograms provides more evidence there may be improvement to be had by using the plan-view approach but that in general, this sample has been challenging to resolve in tomography. 

\begin{figure}
    \centering
    \includegraphics[width=0.83\linewidth]{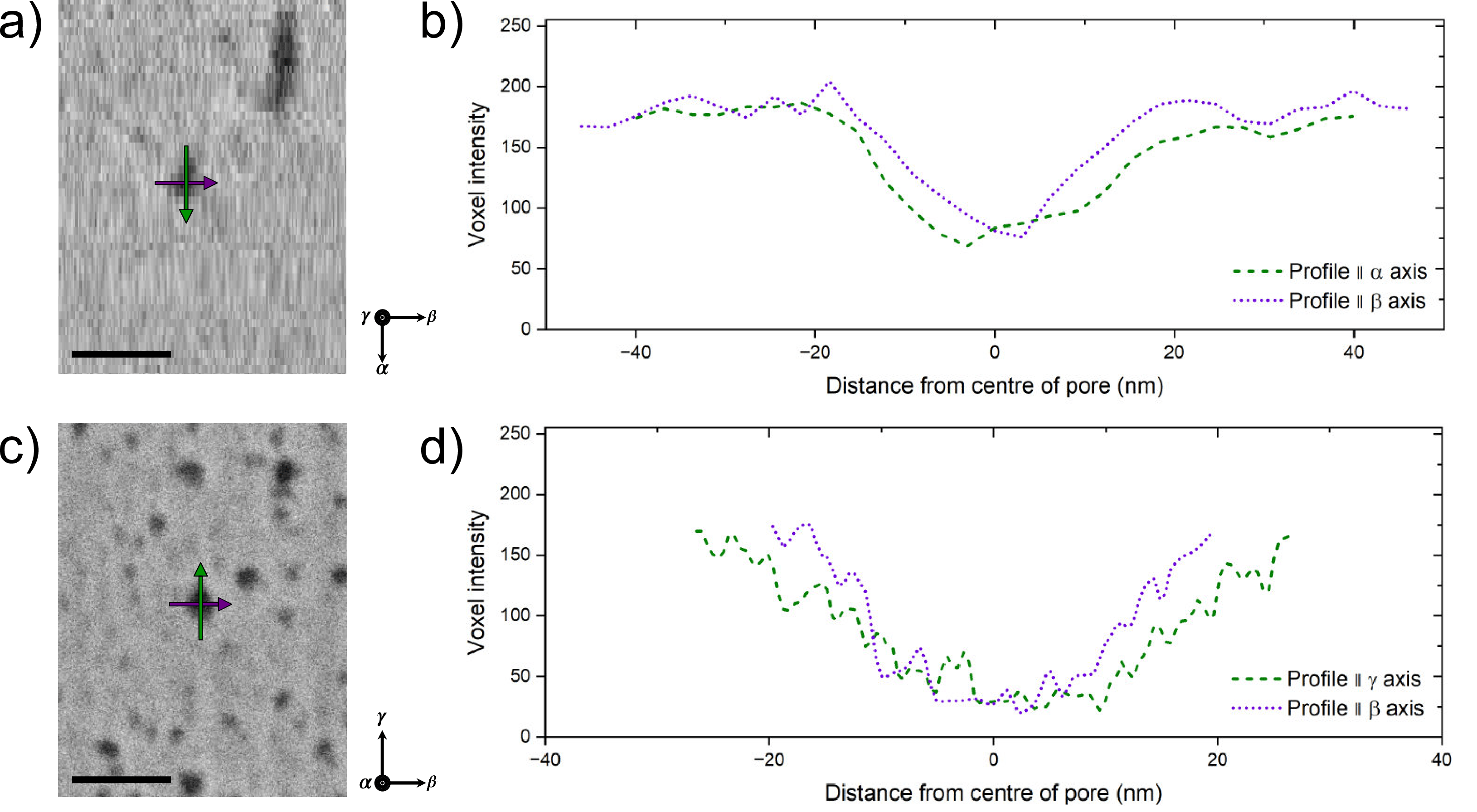}
    \caption{An illustration of feature anisotropy of Sample 2 for the cross-sectional tomograph, showing a) an extracted region of the reconstructed plan-view image from Figure \ref{4VC_XStomo}b highlighted with the position and direction of line profiles and b) the extracted line profiles, compared to the equivalent as-taken images from the plan-view tomograph shown as c) an extracted region of the as-taken plan-view image from Figure \ref{4VC_Vtomo}b highlighted with the position and direction of line profiles and d) the extracted line profiles. Scale bar represents 100\,nm}
    \label{Lineprofilessample2}
\end{figure}

An example of the line profile assessment described above, as applied to pores in Sample 2, is given in Figure \ref{Lineprofilessample2}, equivalently to that of Sample 1 in Figure \ref{Lineprofilessample1}. In general, there is significantly less clear asymmetry for these line profiles than is shown in Sample 1, though, as discussed above, the pore back effect manifests more significantly as a variation in intensity for different pores in the sample rather than as a variation across the width of the pore, simply because far fewer slices feature any one of the fine needle-like pores in Sample 2. Since there is a lower distance across which the pore is imaged, the variation expected is lower, though in principle with a far smaller slice thickness, this analysis could be applied to a sample of the same pore morphology. The line profiles extracted from the as-taken image are roughly equivalent and don't show a great deal of difference between each other or as compared to those extracted from the reconstructed plan-view.

\subsection{Sample 3}
\label{8V oxalic}

Given the success that the plan-view tomography has in resolving pores that are perpendicular to the film surface, it is worth assessing the extent of success for pores that do not follow this morphological form. Given how electrochemical etching takes place on thick doped samples, with pores nucleating on the surface and progressing downwards through the GaN, it is rare to find situations in which pores are entirely parallel to the film surface (where, by symmetry, the cross-sectional approach would have a clear advantage). However, it is not uncommon to see a significant extent of branching in pores etching with some conditions, where they deflect from the vertical etching trajectory into disordered and complex structures \cite{Jiawizzle, Medjahed}. The etching conditions for Sample 3 were selected to address this point effectively, using a sample which shows a significant extent of branching and fine pore diameters. Since the pores in this scenario are neither vertically nor horizontally aligned, neither tomography setup will have an obvious advantage since pores have no clear longest dimension with which to align the milling direction. 

\begin{figure}
    \centering
    \includegraphics[width=1\linewidth]{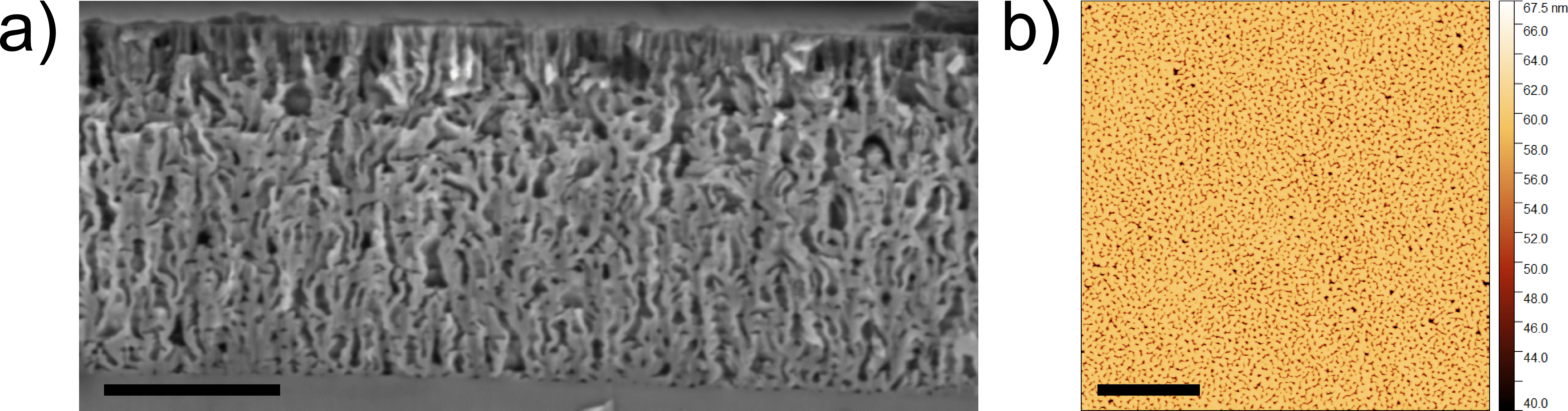}
    \caption{Reference images for the structure of pores in Sample 3, showing a) cross-sectional SEM of a cleaved edge of the porous layer and b) AFM of the surface of the porous layer. Scale bars represent 500\,nm.} 
    \label{sample3Ref}
\end{figure}

Sample 3 was etched in a 0.25\,M aqueous solution of H$_2$C$_2$O$_4$ at a potential of 8\,V in a 2-electrode setup. The structure here is that the pores nucleate on the surface but experience a very high rate of branching as they etch downward through the layer; there is also not always a clear central pipe around which the branching happens and the tortuous structure can expand very far laterally across a huge number of branches. They also have a somewhat different structure at the surface, with a similar fine pore network to that of Sample 1, but connected by a continuous network of surface pores. Once again, Figure \ref{sample3Ref}a gives an SE image of a cleaved cross-section and Figure \ref{sample3Ref}b gives the AFM scan of the surface. Again, the entirety of both tomographs captured on this sample are provided in full in the supplementary material, as `Supplementary Video 3 - Sample 3', which shows four different perspectives analogously to Supplementary Videos 1 and 2, as described in section \ref{8V carbonate}. For both tomographs, the image pixel size was 1\,nm again, and the average slice thickness was 5.8\,nm for the cross-sectional tomograph and 5.9\,nm for the plan-view tomograph.

\begin{figure}
    \centering
    \includegraphics[width=1\linewidth]{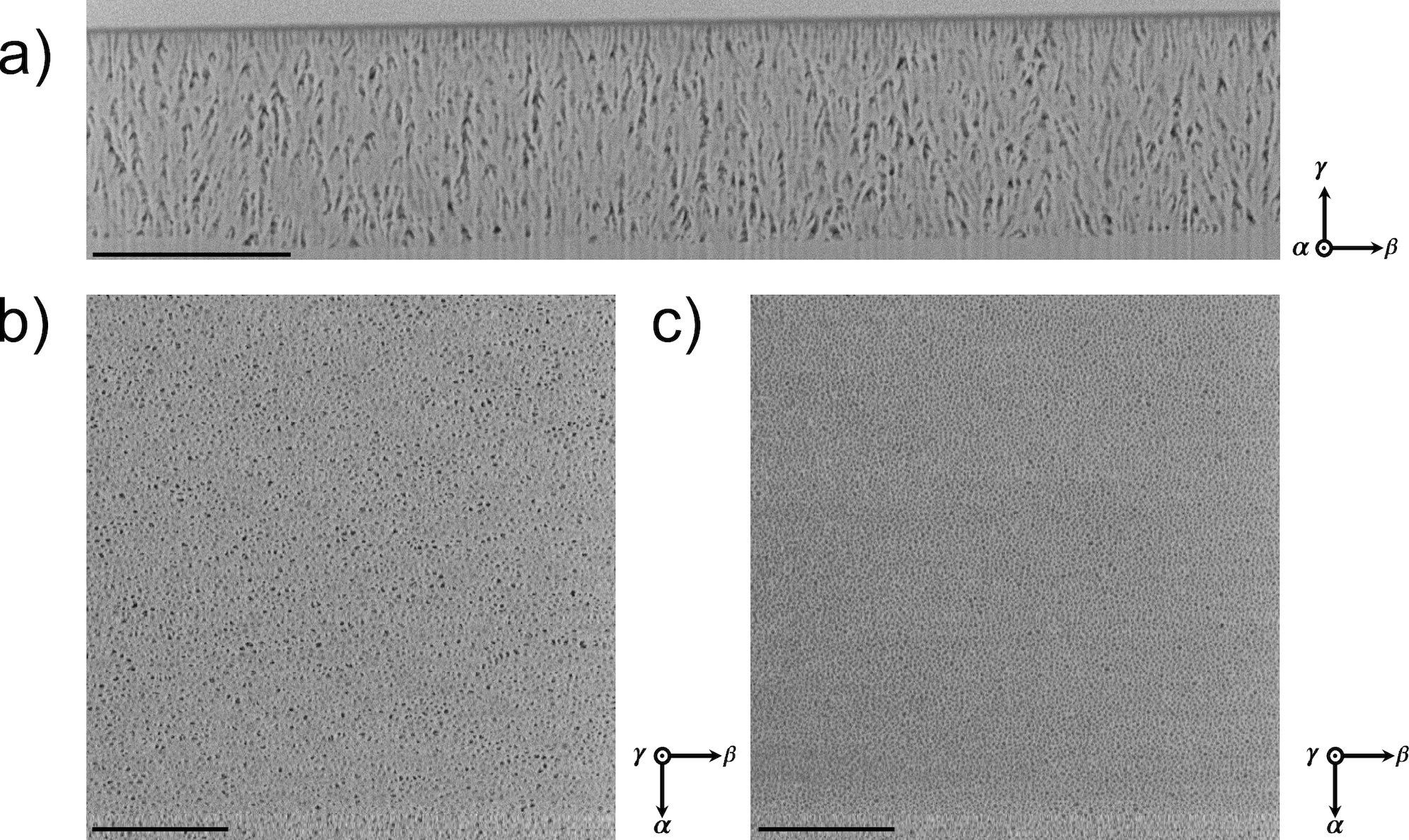}
    \caption{Summarised results from cross-sectional tomography of Sample 3, showing a) an example of an input frame, b) a reconstructed plan-view virtual image positioned at a depth along the $\alpha$ axis of approximately 500\,nm and c) a reconstructed plan-view virtual image positioned at the surface of the porous layer. All scale bars represent 1 \textmu m on the bottom-left of the image and all images have related axes on the bottom-right side.}
    \label{8VOx_XStomo}
\end{figure}

Results of cross-sectional tomography are summarised in Figure \ref{8VOx_XStomo}, in the same layout as Figure \ref{8VC_Vtomo} and \ref{4VC_XStomo}. Figure \ref{8VOx_XStomo}a shows an example of an input frame prepared by milling, which shows a fairly striking difference to the cleaved cross-sectional image of the same sample in Figure \ref{sample3Ref}a. This is easily accounted for by the fact that, since the pores have significant lateral extent, the fracture surface will not be planar, which is even evident from across the top layer of the the sample in the image. With this in mind, the same features can be discerned in both images, showing the branched pores with textured walls and different surface morphology. An example of a reconstructed plan-view virtual image is given in Figure \ref{8VOx_XStomo}b, showing the expected form of a large network of fine pores. When the whole structure is viewed from this perspective, as can be seen in Supplementary Video 3, these pores do not continue vertically down but bifurcate and displace as they move down the layer. The size of these pore features is slightly larger than those in Sample 2 (in Figure \ref{4VC_XStomo}b) and they fill the space more completely. Finally, the reconstructed plan-view virtual image at the surface, given in Figure \ref{8VOx_XStomo}c, shows a very similar pore morphology to the AFM scan in Figure \ref{sample3Ref}b, but with weak contrast to the surrounding GaN due to the small feature size.

\begin{figure}
    \centering
    \includegraphics[width=0.9\linewidth]{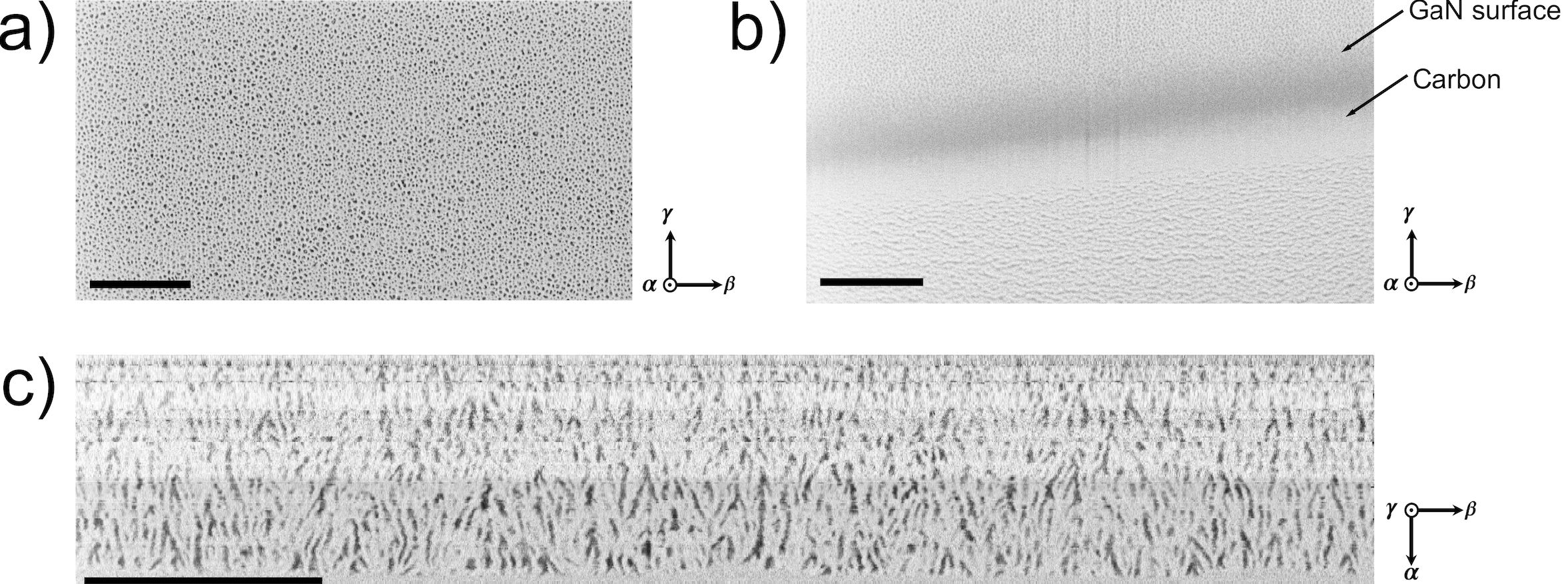}
    \caption{Summarised results from plan-view tomography of Sample 3, showing a) an example of an input frame, b) an early input frame featuring some of the surface of the porous layer, c) a reconstructed cross-sectional virtual image positioned at the centre of the porous layer. All scale bars represent 1 \textmu m on the bottom-left of the image and all images have related axes on the bottom-right side.}
    \label{8VOx_Vtomo}
\end{figure}

Results of plan-view tomography of Sample 3 are given in Figure \ref{8VOx_Vtomo}, again laid out analogously to Figure \ref{8VC_Vtomo} and Figure \ref{4VC_Vtomo}. Figure \ref{8VOx_Vtomo}a shows a typical frame, with intersected pores appearing as dark spots. Figure \ref{8VOx_Vtomo}b shows an example of a frame intersecting the surface, once again with the porous GaN having been mounted on an overhang with the top of the image showing the porous GaN surface and the bottom showing remaining deposited carbon. There are also some curtaining lines from the features in the tracking structure as discussed in section \ref{8V carbonate}. Finally, Figure \ref{8VOx_Vtomo} gives a reconstructed cross-sectional virtual image of the full porous layer. This has been subject to a similar glitching issue to Figure \ref{4VC_Vtomo}c, with the top half experiencing higher intensity and more periodic variation, which is unfortunate, but the features are discernible and show reasonably continuity through the image.

It is a difficult question to compare the two tomographs on how well their reconstructed images conform to the expected morphology for this sample. Directly comparing the tomographs mutually is potentially a more useful exercise than involving the cross-sectional SE image of the cleaved edge in Figure \ref{sample3Ref} since, as mentioned, the cleaving surface is non-planar. The contrast between the pores and GaN for the reconstructed image from the plan-view tomograph in Figure \ref{8VC_Vtomo}c, particularly on the bottom half, is significantly sharper than the contrast on the reconstructed image from the cross-sectional tomograph in Figure \ref{4VC_XStomo}b (indeed, this is because the contrast is also sharper than it is in the input frames for this tomograph, such as in Figure \ref{4VC_XStomo}a, which implies that the pores must have generally more vertical displacement than lateral displacement, which is also evident in the tomography data). With that in mind, it might be suggested that the plan-view tomograph succeeds marginally better, but this is contentious. However, it is worth mentioning that, in general, the reconstructing power of the plan-view tomograph for Sample 3 is probably the weakest of the three plan-view tomographs in this work, and that some of the pore features, particularly in the top half of the image, have very poor contrast from the surrounding GaN. This sample has possibly pushed the limits of FIB tomography in general, and changes to the imaging conditions may be needed for further improvements.

\begin{figure}
    \centering
    \includegraphics[width=0.7\linewidth]{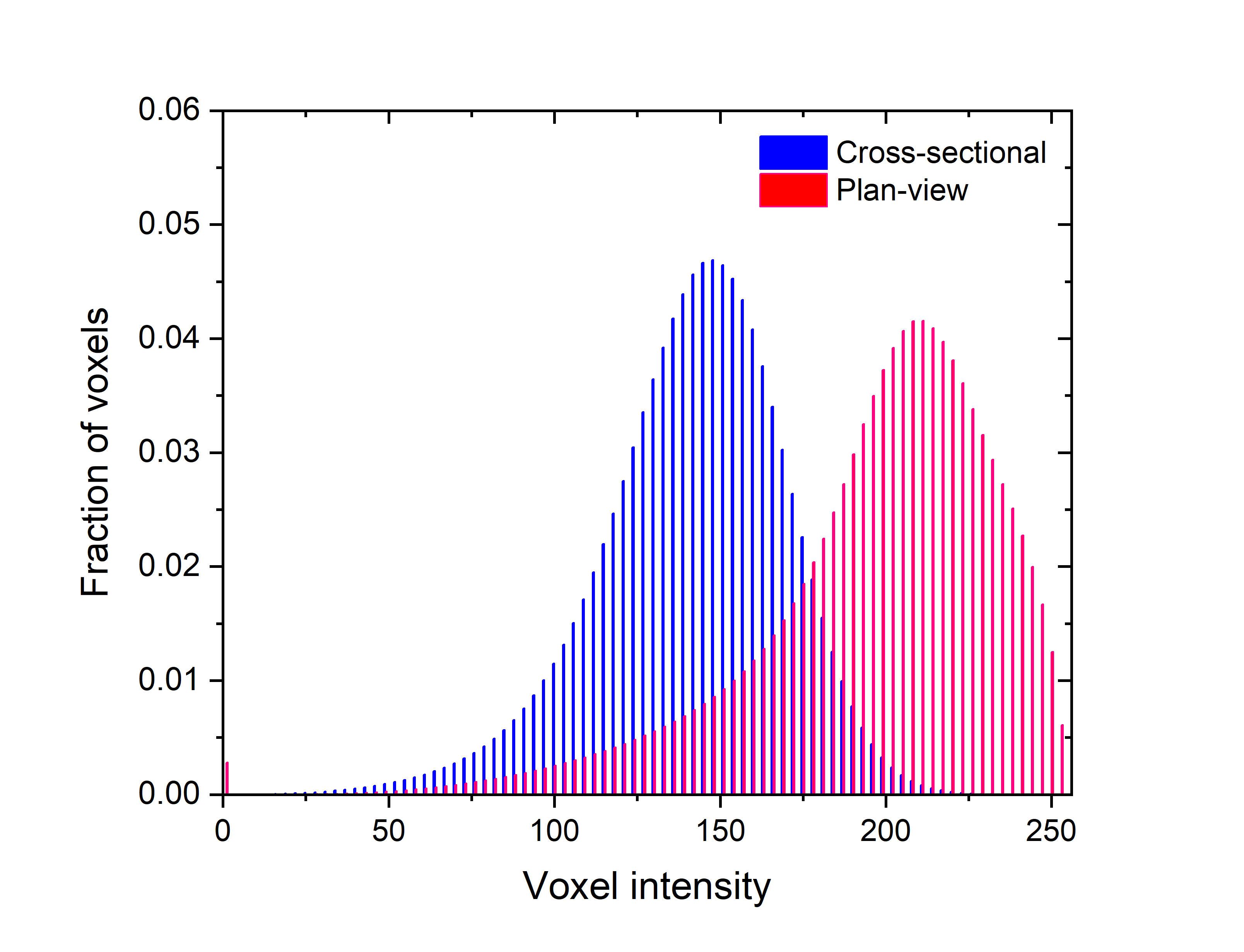}
    \caption{Histograms of distributions of voxel intensity values to compare cross-sectional (blue) and plan-view (red) tomography of Sample 3, showing the value of intensity plotted against the fraction of total voxels in the two tomographs with that intensity}
    \label{sample3Histograms}
\end{figure}

Intensity distributions for the two tomographs on Sample 3 are given in Figure \ref{sample3Histograms}, analogously to Figure \ref{sample1Histograms} and Figure \ref{sample2histograms}. Here, the imaging for the plan-view tomograph has been overexposed, with the distribution capping at 256 without dropping to zero, leading to a loss of feature resolution. This is unfortunate but does not necessarily create issues related to the pore back effect. The two peaks have a similar morphology, and neither show the two-peak effect which would be ideal. However, once again, a very slight shoulder is observed when the data are viewed on a logarithmic scale, as available in Supplementary Figure 3, for the plan-view tomography. There is a very slight change in the gradient at around 130 (voxel intensity) whilst the cross-sectional dataset does not show this as clearly. Once again, this is tenuous, and the histograms show more clearly that both approaches have suffered for the pore back effect in this sample. 

\begin{figure}
    \centering
    \includegraphics[width=0.89\linewidth]{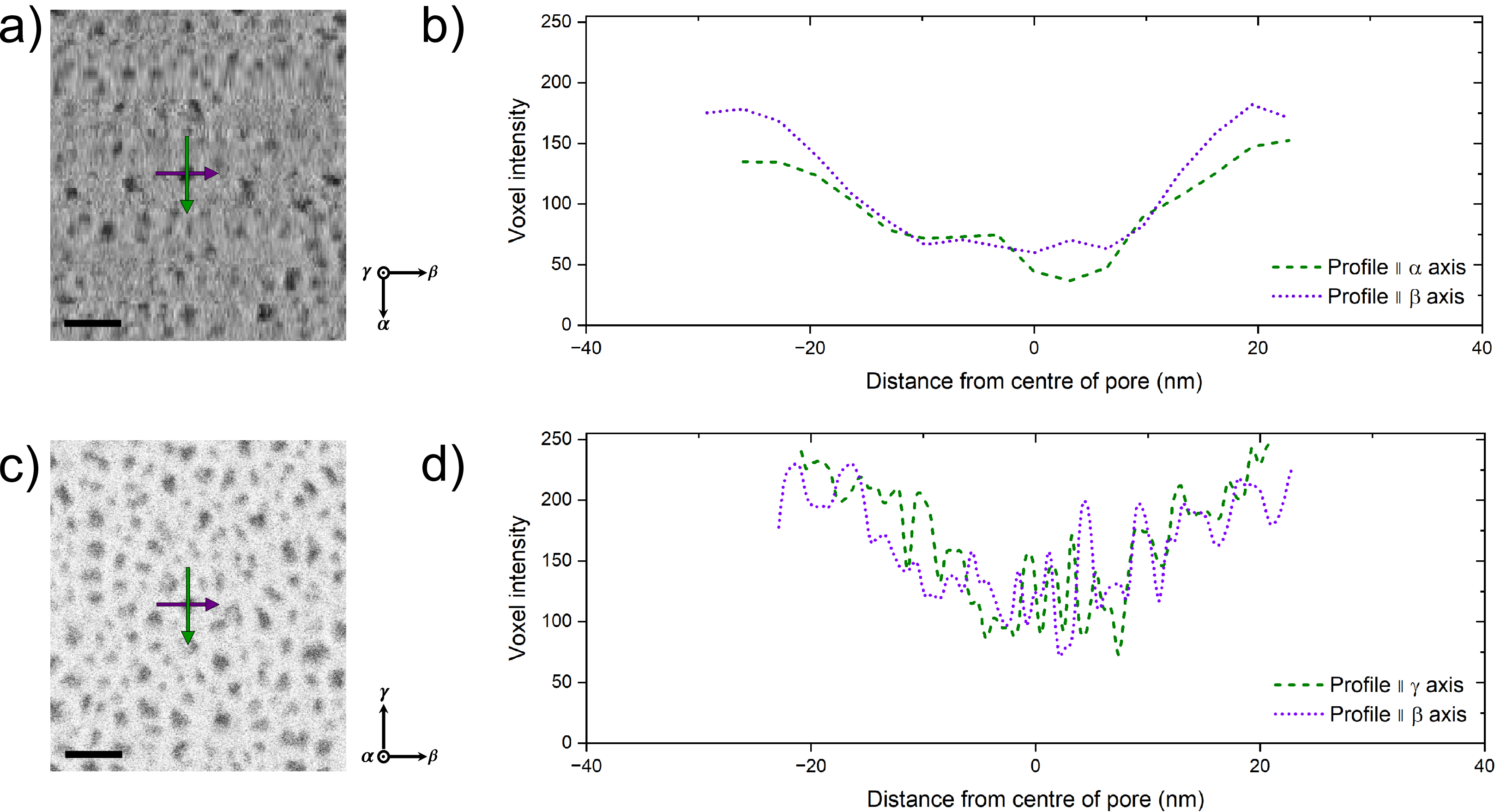}
    \caption{An illustration of feature anisotropy of Sample 3 for the cross-sectional tomograph, showing a) an extracted region of the reconstructed plan-view image from Figure \ref{8VOx_XStomo}b highlighted with the position and direction of line profiles and b) the extracted line profiles, compared to the equivalent as-taken images from the plan-view tomograph shown as c) an extracted region of the as-taken plan-view image from Figure \ref{8VOx_Vtomo}b highlighted with the position and direction of line profiles and d) the extracted line profiles. Scale bar represents 100\,nm}
    \label{Lineprofilessample3}
\end{figure}

Finally, the same line profile treatment applied to Sample 1 and Sample 2 can be applied here for the cross-sectional tomograph. Figure \ref{Lineprofilessample3} shows such a treatment, analogously to Figure \ref{Lineprofilessample1} and Figure \ref{Lineprofilessample2}. Here, we do observe some amount of feature anisotropy along the (positive) $\alpha$ axis once again, as expected. The features in this sample are larger than those in Sample 2 but smaller than those in Sample 1, so this also may give an indication of the minimum feature size for which the method can be used, though this depends on imaging conditions. The pore here shows a symmetrical intensity along the $\beta$ axis and an asymmetrical intensity along the $\alpha$ axis, as expected given the pore back effect. It is worth noting that features in this sample vary fairly significantly in size at different depths, since pore branches terminate at various depths through the layer, and that a larger feature intersecting any given reconstructed plan-view image is necessary to perform the analysis shown. The line profiles extracted from the as-taken image of the plan-view tomograph have quite significant noise and a higher background intensity (for the GaN-equivalent pixels), but they generally present the expected similarity.

\section{Conclusions}
\label{conclusions}

Performing tomography in the plan-view tomography setup for porous c-plane GaN films appears to offer considerably better feature resolution than the conventional approach. This is particularly relevant for pores that propagate vertically downwards through the available doped layer. This new approach may thus be of practical use to those working on characterisation of similar mesoporous thin films, such as electrochemically etched porous alumina \cite{porousalumina} and porous silicon \cite{porousSi}, as well as for porous GaN studies like those shown in this work. However, it is worth noting that the plan-view approach will only remain advantageous for pores of this vertical morphology. We have previously published tomography studies involving porous GaN distributed Bragg reflectors \cite{THORNLEY2026121957, 10.1063/5.0325374}, where the pores form \textit{via} a dislocation-mediated process that leads to a pore morphology in which pores radiated out from a central point within a flat, disc-like region. In this case, the conventional cross-sectional tomography already aligns the milling direction with the long-axis of the pores to the greatest extent possible, and these tomographs have shown good resolving power and feature sharpness. We are now able to achieve such feature sharpness for thicker porous layers, but have to take advantage of anisotropy of features in order to do so. It may therefore be of significant value to develop a simulation workflow, where the structure of pores are 
modelled in three-dimensions using approximate dimensions from other, faster characterisation techniques (such as AFM and SEM of cleaved cross-sections). The capturing of tomograph frames could be simulated by removing slices of given thickness from this geometry that intersect with the pore(s), using Monte Carlo electron scattering or secondary electron generation modelling software. For given pore and voxel dimensions, inputing materials parameters into such a simulation and appraising the resulting datasets using the metrics outlined here could allow for the selection of an appropriate tomograph orientation, but also to model the effect of beam parameters such as electron beam accelerating voltage and optimise those values as well, without the need of running a large series of lengthy tomography experiments on the same sample.

For porous mediums with isotropic pores, there is no preferential orientation for tomography that can be used to mitigate the pore back effect. There are many mesoporous materials that would fall under such a classification. Nonetheless, changing the imaging conditions, such as the electron beam accelerating voltage/probe current, detector and scanning parameters, may lead to some variation in the extent of shine through. Voxel intensity distributions and feature anisotropy measurement techniques have been introduced in this work to appraise the extent of the pore back effect then become valuable for systematic imaging studies, where the effects of the pore back effect can be quantified. Whilst improvements in segmentation algorithms are being made, it is still an important pursuit to ensure that datasets are as reflective as possible of the true structure. The pore back effect has been shown to be damaging to this, hence why studies of this kind will remain crucial for accurate material characterisation. 

\section{Acknowledgements}

This research was supported by the Royal Academy of Engineering under the Chairs in Emerging Technologies Scheme, which is sponsored by the Department for Science, Innovation and Technology (DSIT). Funding was also received from the EPSRC (No. EP/X015300/1). The authors acknowledge the support of Ernest Oppenheimer Trust at the University of Cambridge. We also acknowledge the Royce Institute for the use of the Zeiss Crossbeam 540 under Grant No. EP/R008779/1 and the use of the Bruker Dimension Pro atomic force microscope under Grant Nos. EP/P024947/1 and EP/R00661X/1.

\section{Supplementary Material Available}
\label{SI}

See the supplementary material for the three intensity histograms for the two tomography approaches upon Samples 1-3 given with a logarithmic y-axis in section 1 Also available is extensive discussion of the challenges and development of the setup for the plan-view tomography experiment, in section 2. We also provide videos to show all six tomography datasets in full, with each video composed in the same way; the videos begin with input frames animated in order of their capturing for cross-sectional tomography; the next clip in the video is the reconstructed plan-view images of the cross-sectional tomograph, animated to sweep from the top of the porous layer to the bottom; the third clip in each video is the input frames animated in order of their capturing for plan-view tomography, and the fourth clip is the reconstructed cross-sectional images of the plan-view tomograph, animated to sweep from the left (arbitrary) to right. Supplementary Video 1 shows the datasets in this way for Sample 1, Supplementary Video 2 shows the datasets in this way for Sample 2, and Supplementary Video 3 shows the datasets in this way for Sample 3. Throughout all videos, scale bars represent 1000\,nm and the tomography orientation/image type are labelled in the top left. 

\bibliographystyle{elsarticle-num} \bibliography{reference}

\end{document}